\documentclass[aps,pra,twocolumn,superscriptaddress,showpacs,preprintnumbers,amsmath,amssymb,
longbibliography]{revtex4-2}

\usepackage{epsfig}
\usepackage{epstopdf}
\usepackage{natbib}
\usepackage{amssymb}
\usepackage{soul}
\usepackage[breaklinks=true,colorlinks]{hyperref}
\usepackage{braket}

\allowdisplaybreaks
\begin{document}
\title{Nonlinear coherent heat machines and closed-system thermodynamics}

\author{Tom\'{a}\v{s} Opatrn\'y}
\affiliation{Department of Optics, Faculty of Science, Palack\'y University, 17. listopadu 50, 77146 Olomouc, Czech Republic}
\author{\v{S}imon Br\"{a}uer}
\affiliation{Department of Optics, Faculty of Science, Palack\'y University, 17. listopadu 50, 77146 Olomouc, Czech Republic}
\author{Abraham G. Kofman}
\affiliation{International Center of Quantum Artificial Intelligence for Science and Technology (QuArtist)
and Department of Physics, Shanghai University, 200444 Shanghai, China}
\affiliation{Department of Chemical and Biological Physics,
Weizmann Institute of Science, Rehovot 7610001, Israel}
\author{Avijit Misra}
\affiliation{International Center of Quantum Artificial Intelligence for Science and Technology (QuArtist)
and Department of Physics, Shanghai University, 200444 Shanghai, China}
\affiliation{Department of Chemical and Biological Physics,
Weizmann Institute of Science, Rehovot 7610001, Israel}
\author{Nilakantha Meher}
\affiliation{International Center of Quantum Artificial Intelligence for Science and Technology (QuArtist)
and Department of Physics, Shanghai University, 200444 Shanghai, China}
\affiliation{Department of Chemical and Biological Physics,
Weizmann Institute of Science, Rehovot 7610001, Israel}
\author{Ofer Firstenberg}
\affiliation{Physics of Complex Systems, Weizmann Institute of Science, Rehovot 7610001, Israel}
\author{Eilon Poem}
\affiliation{Physics of Complex Systems, Weizmann Institute of Science, Rehovot 7610001, Israel}
\author{Gershon Kurizki}
\affiliation{Department of Chemical and Biological Physics,
Weizmann Institute of Science, Rehovot 7610001, Israel}

	\begin{abstract}
All existing heat machines are dissipative open systems. Hence, they cannot operate fully coherently.  We propose to replace this conventional thermodynamic paradigm  by  a completely different one,  whereby  heat machines are nonlinear coherent closed systems comprised of few field modes. Their thermal-state input  is transformed by nonlinear interactions into non-thermal output with controlled quantum fluctuations and the capacity to deliver work in a chosen mode. This new paradigm  allows the bridging  of quantum coherent and thermodynamic descriptions.
\end{abstract}
	
\date{\today}
\maketitle
\section{Introduction} 
A heat engine may be viewed as a device capable of two functionalities. First, it {\it must concentrate the energy} of a heat bath in a selected degree of freedom called the working mode, within limits dictated by the first and second  laws of thermodynamics. The enormous number of bath modes characteristic of existing heat engines (HE) justifies the thermodynamic paradigm, dating back to Carnot, that describes HE as {\it open systems dissipated by thermal baths} \cite{callenbook,KondepudiPrigogine,Kosloff_2013,BinderBook,David2015AdvAtMolOpt, Kurizkibook}. Second, the concentrated {\it heat must be partly converted into work output}, which requires the state of the working mode (piston), be it classical or quantum mechanical  \cite{Kosloff_2013,BinderBook,David2015AdvAtMolOpt, Kurizkibook, PhysRevLett.2.262,Alicki_1979, PhysRevE.86.051105,Scully862, PhysRevX.5.031044, Campisi, FTEE, Gelbwaser_2013_b, PRE2014, Niedenzu_2016, e18070244, PhysRevE.94.062109, Ghosh2017, Ghosh2018, niedenzu18quantum, Dillenschneider_2009, Abah_2014, FTE4, Hardal2015ScRep, PhysRevX.7.031044,Mukhopadhyay2018PRE,CarnotPRE,AREPL,Huber,Deffner} to be \textit{non-passive}, i.e. store ergotropy \cite{Gelbwaser_2013_b, PRE2014, Niedenzu_2016,Allahverdyan2004EPL,Pusz_1978,Lenard1978}. The required non-passivity/ergotropy is achievable either via  external control of the piston in  HE \cite{David2015AdvAtMolOpt, Kurizkibook, PhysRevLett.2.262,  Alicki_1979, PhysRevE.86.051105,Scully862, PhysRevX.5.031044, Campisi, FTEE, Gelbwaser_2013_b, PRE2014, Niedenzu_2016, e18070244, PhysRevE.94.062109, Ghosh2017, Ghosh2018, niedenzu18quantum, Dillenschneider_2009, Abah_2014} or via information readout and feedforward by an observer (Maxwell's Demon) in information engines \cite{lutz2015information,Vid2016,e21010065,LeffRex,OpatrnyPRL21}. Other means of producing ergotropy/ non-passivity in HE fed by bosonic thermal fields have not been explored thus far.

These long-standing principles have not been revised by recent trends to incorporate quantum information methodology into thermodynamics,  known as the resource theory of quantum thermodynamics \cite{Gau-pas,Uttam,GA-prl,Varun,CGTO}.  These trends  have mainly  focused on  {\it linear Gaussian operations} (LIGO)  jointly performed on quantum systems and their thermal bath ancillae. The realization and possible implications of  unitary {\it non-Gaussian operations} (NGO), which have been conceived  in quantum optical and quantum information schemes \cite{QND-PRL,QND1,QND2,QND3,QND4}, are mostly uncharted terrain  in the context of HE (but cf. \cite{Ghosh2017}).

Here we break away from the established open-system thermodynamic description and introduce a new paradigm whereby a HE can be a {\it purely coherent, energy-preserving (autonomous) closed system}. This  paradigm is based on {\it nonlinear coupling of thermal bosonic field modes}, which makes  such devices fundamentally different from existing HE that are energized by baths comprised of {\it linearly coupled modes} \cite{callenbook,KondepudiPrigogine,Kosloff_2013, BinderBook, David2015AdvAtMolOpt,Kurizkibook,  PhysRevLett.2.262,  Alicki_1979, PhysRevE.86.051105,Scully862, PhysRevX.5.031044, Campisi, FTEE, Gelbwaser_2013_b, PRE2014, Niedenzu_2016, e18070244, PhysRevE.94.062109, Ghosh2017, Ghosh2018, niedenzu18quantum, Dillenschneider_2009, Abah_2014, FTE4, Hardal2015ScRep, PhysRevX.7.031044,PhysRevLett.109.090601,Gelbwaser_2013_a,Ghosh2018,FTEE,Karimi_2017,Tahir, PhysRevE.99.042121,PhysRevX.5.031044,PhysRevLett.122.110601,PhysRevLett.73.58,Clements:16,PhysRevLett.110.130406,PhysRevE.90.042128,PhysRevLett.110.130406,PhysRevE.90.042128,OpatrnyPRL21,Mukhopadhyay2018PRE,CarnotPRE,AREPL,Huber,Deffner}.  These devices, dubbed here {\it heat engines via nonlinear interference} (HENLI), are described by NGO that, thanks to their hitherto unexploited nonlinearity,  can make the output field modes {\it interfere constructively or destructively}, depending on the nonlinear couplings and the {\it mean quanta numbers} of the thermal input modes. In contrast, LIGO are incapable of performing this feat. The envisaged NGO can  achieve both HE functionalities: heat concentration/steering  from input modes to a designated output mode and partial heat-input  conversion into ergotropy/work output in this mode.

From an information-theoretic perspective, such  NGO cause information flow among the modes, resulting in {\it autonomous} feedforward of the information, as opposed to Maxwell-Demon based information  engines \cite{lutz2015information,Vid2016,e21010065,LeffRex,OpatrnyPRL21}  or externally-controlled HE \cite{Kosloff_2013, BinderBook, David2015AdvAtMolOpt, Kurizkibook, PhysRevLett.2.262,  Alicki_1979, PhysRevE.86.051105,Scully862, PhysRevX.5.031044, Campisi, FTEE, Gelbwaser_2013_b, PRE2014, Niedenzu_2016, e18070244, PhysRevE.94.062109, Ghosh2017, Ghosh2018, niedenzu18quantum, Dillenschneider_2009, Abah_2014, FTE4, Hardal2015ScRep, PhysRevX.7.031044,Mukhopadhyay2018PRE,CarnotPRE,AREPL,Huber,Deffner,Allahverdyan2004EPL,Pusz_1978,Lenard1978}. The information flow and feedforward can change the thermal input states into {\it non-passive (ergotropy-carrying)} output states, while keeping the overall entropy constant (consistently with global coherence/unitarity). 

When viewed quantum-mechanically, these  NGO correlate the thermal input modes in ways inaccessible by LIGO, as quantified by  the nonlinear unitary transformations of their mode-pair Stokes operators \cite{schwinger,agarwalbook,Ulf} and changes of their quantum statistical distributions. When viewed classically, the nonlinear couplings correlate the phases and amplitudes of different modes and thereby {\it stabilize} the interference against the random (thermal) phase and amplitude fluctuations.   Namely, these nonlinear correlations permit to choose parameters that result in predominantly constructive interference of heat in a desired output mode, and destructive interference in undesired output modes.   Significantly, {\it quantum correlations of the coupled modes incur vacuum noise that is a disadvantage for the functioning of HENLI compared to classical nonlinear correlations}.

To gain insight into the proposed  general principle, we assume  hot and cold baths that consist of a discrete set of $N$ modes  taken, for simplicity, to be at the same frequency (the consideration of non-degenerate modes is straightforward but laborious).  These baths can be visualized  as  (ordered or disordered)   {\it mode networks}  that  are divisible into repeated  few-mode blocks: in each block,   few output modes nonlinearly  couple with similar numbers of thermal input modes, instead of (essentially infinite)  mode continua in conventional baths.  Each such block  is a multiport  nonlinear interferometer that is amenable to exact Hamiltonian analysis, both classically and quantum mechanically, without resorting to open-system approaches.

We discuss the HENLI principles from information-theoretic, classical  and quantum mechanical perspectives, first generally and  qualitatively (Sec. II) and then in detail for its  minimal (4-mode)  version (Sec. III) that can serve as a building block of such devices. Mutual  information (MI) on phase-intensity intermode correlation, the  feedforward of this information induced by nonlinear interference and the underlying quantum intermode correlations provide fundamentally new insights  into this formidable problem. We  then extend the analysis  to a cascade of 4-mode HENLI blocks  for scaled-up operation (Sec. IV).

Nonlinear  networks  for HENLI may  be realized in various media, such as phononic structures \cite{QND-PRL} with  anharmonically-coupled modes. Here (Sec. V) we discuss optomechanical setups with nonlinearly coupled photonic and phononic modes \cite{Hekkila2014PRL,GKPNAS}, or photonic modes correlated by cross-Kerr polaritonic interaction \cite{QND1,QND2,QND3,QND4,Shahmoon2011,Friedler,Firstenberg2013,Thompson2017,Tiarks2019}.

 The new paradigm presented here, whereby heat machines can be purely coherent few-mode interferometers, lays the ground  for bridging the conceptual gulf  between nonlinear coherent dynamics and thermodynamics (Sec. VI).


\begin{figure*}
 \includegraphics[width=0.99\linewidth]{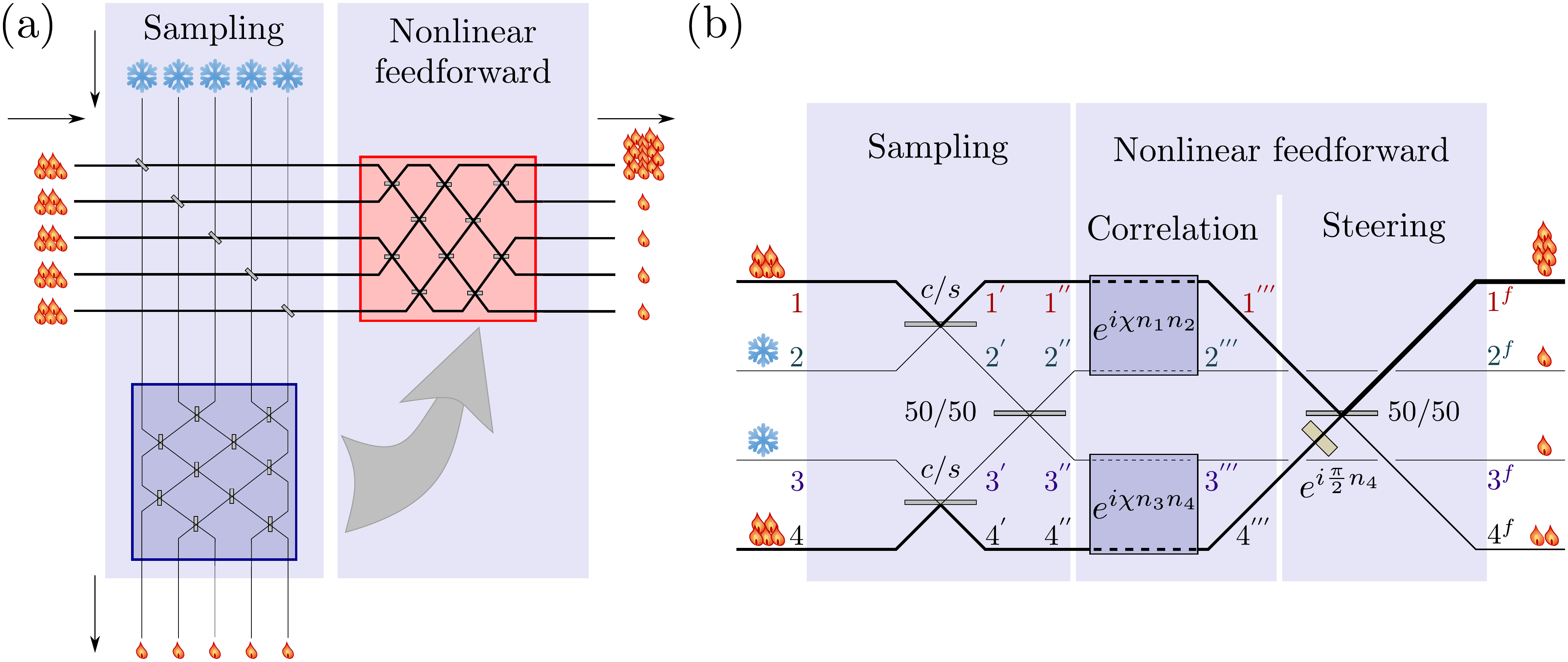}
\caption{\label{Fig1}
 (a) \textbf{A multiport nonlinear interferometer as a heat machine:} Input modes in a hot thermal state and those in a cold state undergo the following processes: (i) A fraction of the energy of the hot modes is split off by beam splitters. The split-off field (in the lower blue box) carries information about (“samples”) the field that remains in the hot modes. (ii) Nonlinear interactions correlate the two boxes. The information stored in the split-off field is converted via this correlation into autonomous feedforward (without the need to read and process the information) such that (iii) the thermal fields can be steered to interfere constructively in one preselected mode by parameter choice. (b)
\textbf{A four-mode heat engine block based on a nonlinear-interferometer (HENLI)} in which the input modes (1 and 4$-$hot, 2 and 3$-$cold) undergo the stages described in the text.  The parameters are chosen such that when averaged over all phases and amplitudes of the hot input modes and adding a linear $\pi/2$ phase-shifter before the final BS, the interference is predominantly constructive in mode $1^f$ and destructive in mode $4^f$.
}
\end{figure*}
\section{HENLI principles}\label{NIHM principles} 
To better understand the need for nonlinear NGO, consider first a linear interferometric network that transforms each input field mode into a linear combination of all others at the output \cite{PhysRevLett.73.58,Clements:16}. The output mode energies then depend on the phase differences of the input field modes. Yet, these phase differences are {\it unknown (random)} if the input is thermal. This randomness prevents selected-mode amplification or heat-to-work conversion \cite{PhysRevLett.110.130406,PhysRevE.90.042128}, because these functionalities are enabled by intermode  mutual-information changes that are absent in coherent LIGO.

Concretely, consider a multiport interferometer with $m$ input modes and $m$ output modes that contains only (energy-conserving) linear mode couplers or beam splitters (BS). If a multimode factorized coherent state $|\beta_1\rangle |\beta_2\rangle \dots |\beta_m\rangle$ is the input, one can find parameters of the interferometer  that give rise to a coherent state $|\alpha\rangle$  in output mode 1, all the remaining output modes being empty, i.e. full energy concentration is achievable. These LIGO are invertible, i.e. for a coherent state $|\alpha\rangle$ at the input of mode 1 with the remaining $m-1$ input modes empty, the output  modes  will  form  a  multimode factorized coherent  state, $|\beta_1\rangle |\beta_2\rangle \dots |\beta_m\rangle$.

If, however, the input is thermal noise, it can be treated as a distribution of coherent states $|\beta_1\rangle |\beta_2\rangle \dots |\beta_m\rangle$ with random amplitudes of $\beta_1, \beta_2, \dots \beta_m$ that have Gaussian distributions with zero mean.  Neither of the HE functionalities are then achievable: (i) The  multimode interference is washed out, prohibiting energy concentration. Notably, if the input modes have equal temperature (mean intensity), so will the output modes under LIGO. (ii) A passive (thermal state) in each input mode remains passive in the corresponding output mode, i.e.  no  work  can  be  extracted from the output by single-mode unitaries, since LIGO cannot change the state passivity \cite{Gelbwaser_2013_b, PRE2014, Niedenzu_2016,Allahverdyan2004EPL,Pusz_1978,Lenard1978}.

Yet, if we could estimate the magnitudes and phases of $\beta_1, \beta_2, \dots \beta_m$, and feedforward the results for each realization of the random input, we would be able to choose the interferometer parameters such that the energy {\it is mostly concentrated} in mode 1 via constructive interference and the output state in this mode is rendered {\it non-passive} via NGO. Here we show that it is possible to achieve these estimation and feedforward {\it autonomously}  by nonlinear cross-couplers that correlate mode pairs via their intensity-dependent phase shifts. 

 Such  autonomous  intensity and phase  estimation cannot be achieved if all the input modes are in the same thermal state: Part of the inputs have to be (desirably) empty or (at least) in a colder thermal state than the hot ones, and thus serve as the cold (few-mode) bath. Otherwise, the second law, which prohibits the operation of HE fed by a single-temperature source, would be violated.

 The general $m$-mode protocol  of  HENLI  consists of two major stages (see Fig. 1 (a)):
 
    a) {\it Sampling}: A fraction of each hot input field mode is split off and mixed with a  corresponding cold field mode  by a LIGO which  is described  as field-quadrature rotation in $m$-mode space, realized in optics by imbalanced BS.  The  cold-mode states then become {\it weak copies} of the respective hot-mode states: perfect weak copies if the cold modes are empty, imperfect if they are not.  These $m/2$ copies are combined by LIGO rotations, optically realized by 50/50 BS for pairs of weak copies. These LIGO will be shown to ``sample'' the phase-difference distribution of the input-modes:  the ``sampling'' results are stored (encoded) by the intensity mixing ratio of the weak-copy outputs.  
    
b) {\it Nonlinear feedforward}: As we show, only if  these weak-copy  outputs are nonlinearly cross-correlated  by NGO (here chosen to be nonlinear Kerr  cross-coupling)  with the set of the dominant  (strong) hot-mode fractions, can the mixing ratio of the hot output modes at the final BS be controlled by the nonlinearity.  These  NGO can render the output distributions \textit{non-thermal}. An additional LIGO ($m$-mode basis rotation and phase-shifting) exploits this nonlinear feedforward of the sampling to steer/concentrate the energy mainly to the desired mode. 
 
These operations will be investigated in Sec. III for the minimal (4-mode) HENLI version. The analysis holds, however,  for {\it any size of an  interferometric} network which can be divided into 4-mode blocks (Sec. IV).

\section{Minimal HENLI analysis}
Energy concentration must involve at least two  hot modes, and each such mode is sampled (copied) by at least one cold mode (\textit{the more copies the better}). Therefore, the minimal version of HENLI (shown in Fig. 1(b)) contains {\it two
  hot and two cold input modes at the same frequency}, labeled 1, 2, 3, and 4 from top to bottom. Extension to modes of different frequencies, to be presented elsewhere, does not reveal essentially new insights. Henceforth, we assume for simplicity that the cold input modes 2 and 3 are empty (an assumption that can be relaxed).   Another assumption is that the coherence length of this 4-mode interferometer is much longer than its spatial size, so that temporal evolution can be replaced by {\it discrete steps} (stages), each described by  a unitary evolution operator.

\subsection{Phase-intensity  distribution  and information}
   We seek to control the energy transfer between the hot  modes 1 and 4, and  their output work capacity (ergotropy or non-passivity \cite{Gelbwaser_2013_b, PRE2014, Niedenzu_2016,Allahverdyan2004EPL,Pusz_1978,Lenard1978}). To this end, we estimate and feedforward the apriori unknown amplitudes of the hot modes 1 and 4 so as to let them interfere, predominantly constructively  in one port of the output BS and destructively in the other port. Instead of conventional measurements that can provide such information nearly perfectly \cite{OpatrnyPRL21}, we settle here for partial, imperfect information that is autonomously extracted from cross-Kerr nonlinear coupling. It is extracted in the form of {\it mutual information (MI)} of the quantum-number difference $n\_$ and phase difference $\phi$ of between modes 1 and 4. 

\begin{figure*}
 \includegraphics[width=0.95\linewidth]{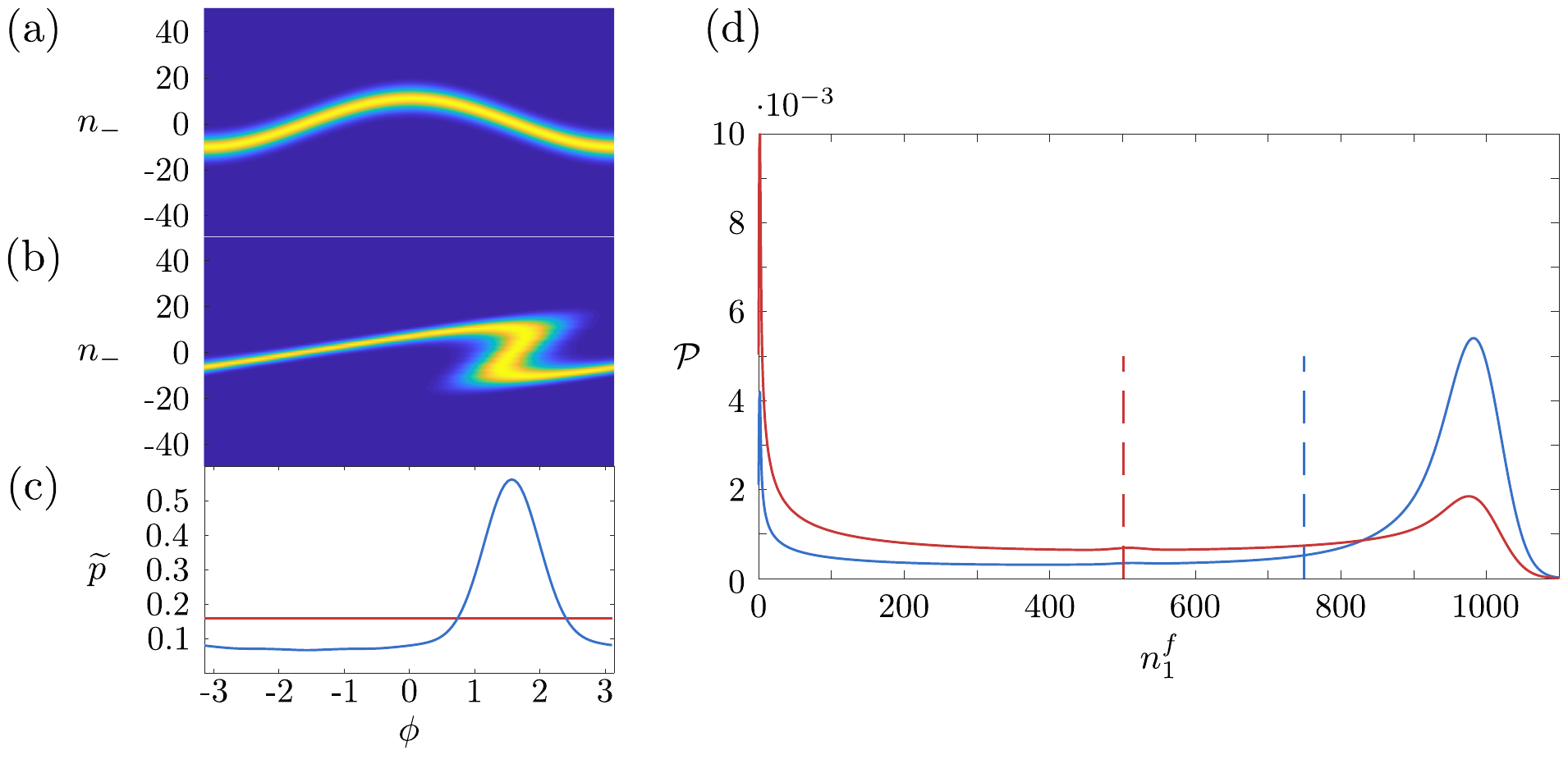}
\caption{\label{Fig2}
 (a) \textbf{Joint distribution $\mathcal{P}(n_-,\phi)$ of $n_-$ and $\phi$ after sampling stage.} Brighter color denotes higher probability. (b) Same distribution after cross-Kerr stage. The change in $n_--\phi$ correlations yields a change in their mutual information.
(c) Phase-difference distribution$-$ initial (red) and final (blue). 
(d) Output quanta-number distribution in mode $1^f$ for input thermal distribution. Red$-$no cross-Kerr, blue$-$with  cross-Kerr, output with constructive interference (+ sign in Eq. (\ref{Outputmean})).
}
\end{figure*}

In each ($k$-th)  stage  of HENLI, their evolving correlations determine the MI, denoted by $I_k$, in terms of the joint distribution $\mathcal{P}_k(n\_, \phi)$  and the marginal distributions $p_k(n\_)$, $\tilde{p}_k( \phi)$ \cite{parrondo2015thermodynamics}. The MI $I_k$ usable for feedforward is given by \cite{Cover}
\begin{equation}\label{MI}
 I_k=\sum_{n\_}\int_0^{2\pi}\mathcal{P}_k(n\_, \phi) \ln\frac{\mathcal{P}_k(n\_, \phi)}{p_k(n\_)\tilde{p}_k( \phi)} d\phi.
\end{equation}
Initially, the $n_-$ and $\phi$ distributions are uncorrelated, $\mathcal{P}_0(n\_, \phi)=p_0(n\_)\tilde{p}_0(\phi)$, where $\tilde{p}_0(\phi)=1/(2\pi)$ is the uniform distribution, so that their MI is zero.
Total entropy conservation due to HENLI unitarity implies that $I_k$ in Eq. (\ref{MI}) changes from one stage to another as a function of the entropies \cite{Kurizkibook,Misra}
\begin{align}
 I_k=\mathcal{S}_k(\{n\_\})&+\mathcal{S}_k(\phi)-[\mathcal{S}_{k-1}(\{n\_\})+\mathcal{S}_{k-1}(\phi)],\\
 \mathcal{S}_k(\{n_-\})&=\sum_{\{n\_\}} p_k(n_-)\text{ln}[p_k(n_-)],\nonumber\\
 \mathcal{S}_k(\phi)&=\int_{-\pi}^{\pi} d\phi \tilde{p}_k(\phi)\text{ln}[\tilde{p}_k(\phi)].\nonumber
\end{align}
This MI is evaluated for the consecutive stages of HENLI (Fig. 1(b), left to right):
  
(a) At the  {\it sampling stage},  the first BS, with low  transmissivity $s = \sin \theta \ll 1$  (high reflectivity $c= \cos \theta \lesssim 1$ ) cause small fractions of the hot input modes $1$ and $4$ to be  split off and merge, respectively, with  the empty modes $2$ and $3$. At the output of these BS, we then have weak copies $2'$ and $3'$ of each  coherent-state amplitude in the random thermal distributions of  modes $1$ and $4$, respectively. Namely, for each coherent-state realization, these weak copies have  intensity  difference that is proportional to $n\_$ and the same phase-difference $\phi$ as the input modes 1 and 4. The weak copies then merge on a $50/50$ BS whose  output modes $2''$ and $3''$  have amplitudes  determined by the unknown $-\pi \leq \phi\leq \pi$. This is the ``sampling'' stage, since the initially uncorrelated distributions of $n\_$ and $\phi$ acquire correlations  at the output of the $50/50$ merger. 
 When averaged over the thermal input ensembles (see SI), these correlations  give rise to MI on the weak-copy modes (2-3) that encodes $n\_$  via its dependence on $\cos \phi$ at the output of 50/50 merger. This MI has the same form as Eq. (\ref{MI}). As a result, in the weak-copies the $n\_$ distribution  broadens (increases its entropy) while the $\phi$ distribution is still uniform, as shown in Fig. \ref{Fig2}a.
 
(b) {\it At the  nonlinear feedforward stage}, two Kerr cross-couplers cause, in the classical approximation \cite{Boyd}, the phase of the  hot mode $1''$  to  be  shifted  proportionally  to the  intensity  of mode $2''$ and  the  phase  of  the  hot  mode $4''$   to  the  mode $3''$ intensity, so that  $\phi\rightarrow \phi +\chi n\_''$. Thus, the cross-Kerr effect causes the coherent states with  amplitudes $|\alpha_1''\rangle$ and $|\alpha''_{2}\exp(i\phi)\rangle$  to become $|\alpha_1''\exp(i \chi \alpha_2^2)\rangle$ and $|\alpha_2'' \exp(i\phi+i \chi \alpha_1^2)\rangle$, respectively. 

The exact quantum description of HENLI must account for the intermode entanglement of the field states that emerge from the sampling stage (see Sec. IIIB). 
For the MI analysis, however, it suffices to replace the exact intermode quantum correlations  by their mean \textit{in the classical approximation}
which neglects quantum fluctuations and entanglement. For each set of coherent state amplitudes in the thermal input distribution, $\{\ket{\alpha_1},\ket{0_2},\ket{0_3},\ket{\alpha_4 e^{i\phi}}\}$, we then have the following evolution stages:\\
(i) \textit{Nonlinear correlation}: After the cross-Kerr couplers, the coherent-states in the strong-fraction  hot modes $1'''$ and $4'''$ become, in this classical approximation,
\begin{eqnarray}
\label{EaafterKerr1Clas}
 |\alpha_1^{\prime \prime \prime}\rangle 
&=& |c\alpha_1 \exp \left\{i\chi \frac{s^2}{2} \left|\alpha_1+\alpha_4 e^{i\phi}\right|^2 \right\}\rangle, \\ \nonumber
 |\alpha_4^{\prime \prime \prime} \rangle
&=& |c\alpha_4 \exp \left\{i\left[\chi \frac{s^2}{2} \left|\alpha_1-\alpha_4 e^{i\phi}\right|^2 + \phi \right] \right\}\rangle.
\end{eqnarray} 
Thus, the phase shifts of the hot modes $1'''$ and $4'''$ that are merged by the second $50/50$ BS are  {\it nonlinearly correlated} and depend on $\phi$ {\it non-sinusoidally}. We can infer the MI  at the output from the $(n_--\phi)$ distribution of the modes $1'''$ and $4'''$ (Fig. \ref{Fig2}b, SI). {\it The MI at the nonlinear correlation stage}  ($k=2$ in Eq. \ref{MI}) {\it is diminished} compared to that of the sampling stage ($k=1$), i.e. $I_k$ is partly consumed to reduce $\mathcal{S}_k(\phi)$ while $\mathcal{S}_k(n\_)$ is unchanged. This  means that the Kerr cross-coupling {\it feeds forward the information} encoded by $\phi$.
The result of this nonlinear feedforward is a \textit{sharply peaked (narrowed-down)} $\phi$ distribution (Fig. \ref{Fig2}c). 

(ii) \textit{Steering}: The final 50/50 BS, preceded by a $\pi/2$ shift of mode $4'''$, yields at the two outputs  the coherent-state amplitudes
 $\alpha_{1,4}^{f} = 2^{-1/2}(\alpha_1^{\prime \prime \prime}\pm i \alpha^{\prime \prime \prime})$, which determine the output intensities
\begin{align}
|\alpha^f_{1,4}|^2 = \frac{c^2}{2}\left[
\alpha_1^2 + \alpha_4^2 \pm 2 \alpha_1 \alpha_4 
\sin \left( 2 s^2 \alpha_1 \alpha_4  \chi \cos \phi - \phi \right) \right].
\end{align}
 
This  {\it non-sinusoidal} dependence of the interference term on the phase difference $\phi$ of the input fields, stems from the nonlinearity of the feedforward. The final {\it narrow-peaked} $\tilde{p}^f(\phi)$ allows, for appropriate $\chi$ and $s$, to  achieve  predominantly  destructive interference in mode $4^f$ and constructive interference in mode $1^f$ and thereby net transfer (steering) of mean intensity (energy) from mode $4$ to mode $1$ (or conversely), upon averaging over the random input amplitudes and phase differences  $\phi$ in the thermal input distribution, \mbox{$\mathcal{P}(\alpha_1,\alpha_4,\phi) = \frac{2}{\pi \bar n^2}\alpha_1 \alpha_4 e^{-\frac{\alpha_1^2+\alpha_4^2}{\bar n}}$} (Fig. \ref{Fig2}d).

\begin{figure}
\includegraphics[width=0.85\linewidth]{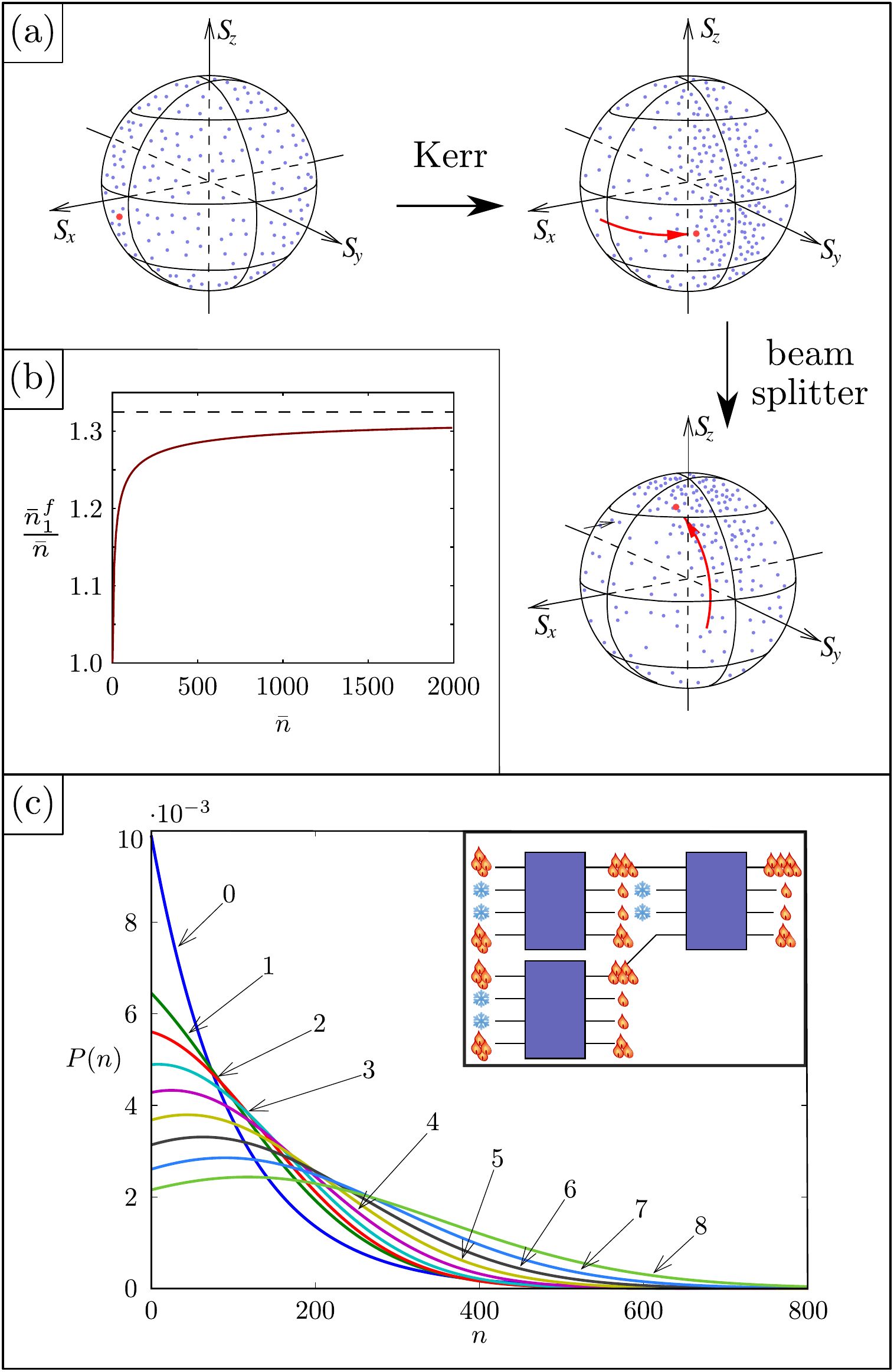}
\caption{\label{f-spheres}
(a) \textbf{Scheme of the transformations on the 2-mode Poincar\'{e} sphere of modes 1 and 4.} The cross-Kerr interaction concentrates the states in the negative $S_y$ hemisphere and the final beam splitter rotates the sphere by $\pi/2$ around $S_x$ to move the states towards $S_z>0$.
Blue dots represent  coherent states initially distributed randomly on the Poncar\'{e} sphere. The red point is a randomly chosen state for which the red arrows show the transformations.
The distorted distribution is eventually concentrated near the north pole (maximal $\bar S_z$).  (b) Mean output quanta number in mode $1^f$, $\bar n_{1}^f$, normalized to the mean input quanta number $\bar n$ plotted vs $\bar{n}$ for thermal states of equal temperature in the hot modes 1 and 4  and optimized parameters $\chi$ and $s$. Full line: fully quantum calculation, broken line: classical approximation (see SI). \textbf{ (c)}  Inset: \textbf{Cascading the HENLI blocks:} The highest-energy outputs of each  block are used as the hot-mode inputs of the next one, thus gradually increasing the mean energy of a preselected mode.; \textbf{ Quanta-number distribution} of the highest energy outputs in 8 consecutive blocks shows growing displacement of the distribution mean, corresponding to an increasing \textbf{\textit{work capacity (non-passivity)}} of the output.}
\end{figure}

We highlight the case of \textit{equal input temperatures} which clearly shows that without cross-Kerr coupling there is no steering. The mean intensities at the output of the final $50/50$ BS in the strong-fraction hot modes are, in this classical approximation, 
\begin{eqnarray}\label{Outputmean}
\bar{n}^f_{1,4} &=& c^2 \bar n\left[ 1 \pm  \frac{s^2 \chi \bar n}{(1+s^4 \chi^2 \bar n^2)^2} \right] .
\label{nbarThermClas}
\end{eqnarray}
The optimal value of $\chi$, $\chi_{\rm opt} = \frac{1}{\sqrt{3}\bar n s^2}$ when inserted in (\ref{nbarThermClas}), yields $(\bar{n}^f_1)_{(\rm max)} = c^2 \bar n \left( 1+\frac{9}{16\sqrt{3}} \right)\approx (4/3) c^2 \bar n$ (Fig. \ref{f-spheres}(b)).
This result shows that the best strategy would be to split off as little of the incoming energy as possible ($s^2\ll 1$) and allow for large Kerr nonlinearity ($\chi\gg 1$).

The steering ability due to nonlinear cross-coupling comes at a price: Part of the input energy leaks to the cold (initially empty) modes $2$ and $3$, so as to conserve sum of entropies of all modes, as required by the coherence of the process (SI).

\subsection{Intermode quantum correlations}

{\it (i) Stokes operators}: The quantum correlations of the HENLI modes and their output nonthermal states that allow for steering of heat and work can be concisely captured by the two-mode Stokes operators \cite{schwinger,agarwalbook,Ulf,Jauch,Robson}. The Stokes operators  are expressed  for modes $i$ and $j$ in terms  of their annihilation operators $\hat a_{i(j)}$ 

\begin{eqnarray}
&&\hat{\mathcal{J}}_x^{(ij)}=\frac12(\hat{\mathcal{J}}_+^{(ij)}+\hat{\mathcal{J}}_-^{(ij)}),~~
\hat{\mathcal{J}}_y^{(ij)}=\frac{-i}{2}(\hat{\mathcal{J}}_+^{(ij)}-\hat{\mathcal{J}}_-^{(ij)}),
\nonumber\\
&&\hat{\mathcal{J}}_+^{(ij)}
=\hat{a}_i^\dagger \hat{a}_j,\label{16} ~~\hat{\mathcal{J}}_-^{(ij)}
=\hat{a}_i\hat{a}_j^\dagger,\nonumber\\
&&\hat{\mathcal{J}}_z^{(ij)}=\frac12(\hat{a}_i^\dagger\hat{a}_i-
\hat{a}_j^\dagger\hat{a}_j),
~\hat{\mathcal{J}}_0^{(ij)}=\frac12(\hat{a}_i^\dagger \hat{a}_i
+\hat{a}_j^\dagger\hat{a}_j).
\end{eqnarray}

The fully quantum dynamics of the Stokes operators  in  HENLI is governed by the product of multimode  unitary evolution operators that corresponds to the successive sampling (sa), cross-Kerr nonlinear (nl) and steering (st) stages  described above. For the $4$-mode version we have
the unitary operator
\begin{eqnarray}\label{Unitary}
&&\hat{U}=\hat{U}_{\rm st}\hat{U}_{\rm nl}\hat{U}_{\rm sa},\nonumber\\ 
 &&\hat{U}_{\rm sa}=\hat{U}_{\rm bs}^{(23)}(\pi/4)\hat{U}_{\rm bs}^{(12)}(\theta)
\hat{U}_{\rm bs}^{(34)}(\theta),\nonumber\\
&&\hat{U}_{\rm nl}=\hat{U}_{\rm cK}^{(12)}\hat{U}_{\rm cK}^{(34)};~~ \hat{U}_{\rm cK}^{(ij)}=e^{i\chi\hat{n}_i\hat{n}_j},
\nonumber\\
&&\hat{U}_{\rm st}=\hat{U}_{\rm bs}^{(14)}(\pi/4)e^{-i\pi\hat{n}_4/2},
\end{eqnarray}
where $\hat{U}_{\rm bs}^{(ij)}(\theta)=e^{2i\theta \hat J_y^{(ij)}}$ is a two-mode BS-mixing operator and $\hat{U}_{\rm cK}$ is the cross-Kerr operator. Both $\hat{U}_{\rm sa}$ and $\hat{U}_{\rm st}$ are LIGO rotations on the 4-mode Poincar\'e sphere, whereas $\hat{U}_{\rm nl}$ is a ``twisting'' NGO that entangles all modes in a nonlinear fashion.

The Stokes operators at the output ($f$) are found to be related to their input (\textit{in}) counterparts in an exactly solvable, intricate nonlinear fashion (SI). The total population of the four modes is conserved according to the {unitarity of the transformation Eq. (\ref{Unitary}),} while the population difference operator of the hot output modes $1$ and $4$ is determined  by the ($\chi$-dependent) exponential of combined $\hat{\mathcal{J}}_x^{(ij)}$ operators of all four modes (SI). Whereas for $\chi=0$, this is a trivial BS-induced LIGO, 
the population difference operator $(\hat{\mathcal{J}}_z^{(14)})^{(f)}$ for $\chi\neq0$, depends {on the intermode correlations created at the sampling stage} via the exponential factor $e^{-2i \chi \hat{\mathcal{J}}_x^{(23)'}}$, where $\hat{\mathcal{J}}_x^{(23)}{}'=
\left[-s^2\hat{\mathcal{J}}_x^{(14)}+
c^2\hat{\mathcal{J}}_x^{(23)}+
sc(\hat{\mathcal{J}}_x^{(24)}-\hat{\mathcal{J}}_x^{(13)})\right]^{(in)}$. This quantum-nonlinear Kerr-induced NGO distorts (``twists'') and entangles the input inter-mode distribution.

{\it (ii) Stokes parameters} are the mean values of the Stokes operators.
{Here we are interested in the Stokes parameters of modes 1 and 4,
\begin{equation}
 S_k=\braket{\hat{\mathcal{J}}_k^{(14)}},\quad k=x,y,z,0.
\end{equation}}
One can associate any two-mode coherent state with a point $(S_x,S_y,S_z)$ on the surface of a Poincar\'{e} sphere of radius $S_0$. 
Under thermal-ensemble averaging, the first two BS are (``sampling'') LIGO that rotate the distribution, but the crucial nonlinear stage 
{\it concentrates} the random phase-space points on the surface of the hemisphere $S_y<0$ (Fig. \ref{f-spheres}a).
The $\pi/2$ phase shifter in mode 4$^{\prime \prime \prime}$ and the output BS correspond to a rotation around $S_x$ by $\pi/2$. This steering stage brings the phase-space points concentrated on the   $S_y<0$ hemisphere surface to the  $S_z>0$ hemisphere surface.
Such {\it energy steering} does not violate the Liouville theorem, since the contraction of the phase-space volume occupied by modes 1 and 4 is compensated by the expansion of the volume corresponding to modes 2 and 3.

When the input modes 1 and 4 are {quantum coherent states} with equal amplitudes, $\alpha_1=\alpha_4$ and random phases,
their Stokes parameters at the output, averaged over their random phase difference $\phi$, yield {(SI)}
\begin{equation}\label{e2}
\bar{S}_x^{\rm(f)}=\bar{S}_y^{\rm(f)}=0,\quad
\bar{S}_z^{\rm(f)}=c^2\alpha_1^2e^{-d}J_1(b),\quad
\bar{S}_0^{\rm(f)}=c^2\alpha_1^2,
\end{equation}
where {the overbar denotes the average over $\phi$,}  $d=s^2(1-\cos \chi)(\alpha_1^2+\alpha_4^2)$ and $J_1(b)$ is the first order Bessel function with argument $b=2s^2\alpha_1\alpha_4 \sin \chi$.
The average quanta numbers of the output modes are found to be

\begin{equation}\label{e3}
\overline{\braket{\hat{n}_{1,4}^{\rm(f)}}}=
\bar{S}_0^{\rm(f)}\pm\bar{S}_z^{\rm(f)}=
c^2\alpha_1^2[1\pm J_1(b)e^{-d}],
\end{equation}
where $\hat{n}_i=\hat{a}_i^{\dagger}\hat{a}_i$.
When the fields in HENLI are classical, we obtain the average Stokes parameters and field intensities given by Eqs.~(\ref{e2}) and (\ref{e3}) with $d=0$ and $b=2s^2 \alpha_1 \alpha_4 \chi$ (the quantum and classical expressions for $b$ coincide when $\chi\ll 1$ $-$see SI).
A nonzero $d$ results from quantum entanglement, created by the Kerr effect, and vacuum fluctuations of the modes. The exponential decay factor $d$ thus diminishes the energy steering in Eqs. (\ref{e2}), (\ref{e3}) as compared to Eq. (\ref{Outputmean}) in the classical approximation (Fig. \ref{f-spheres}b).

\begin{figure*}
\centerline{\epsfig{file=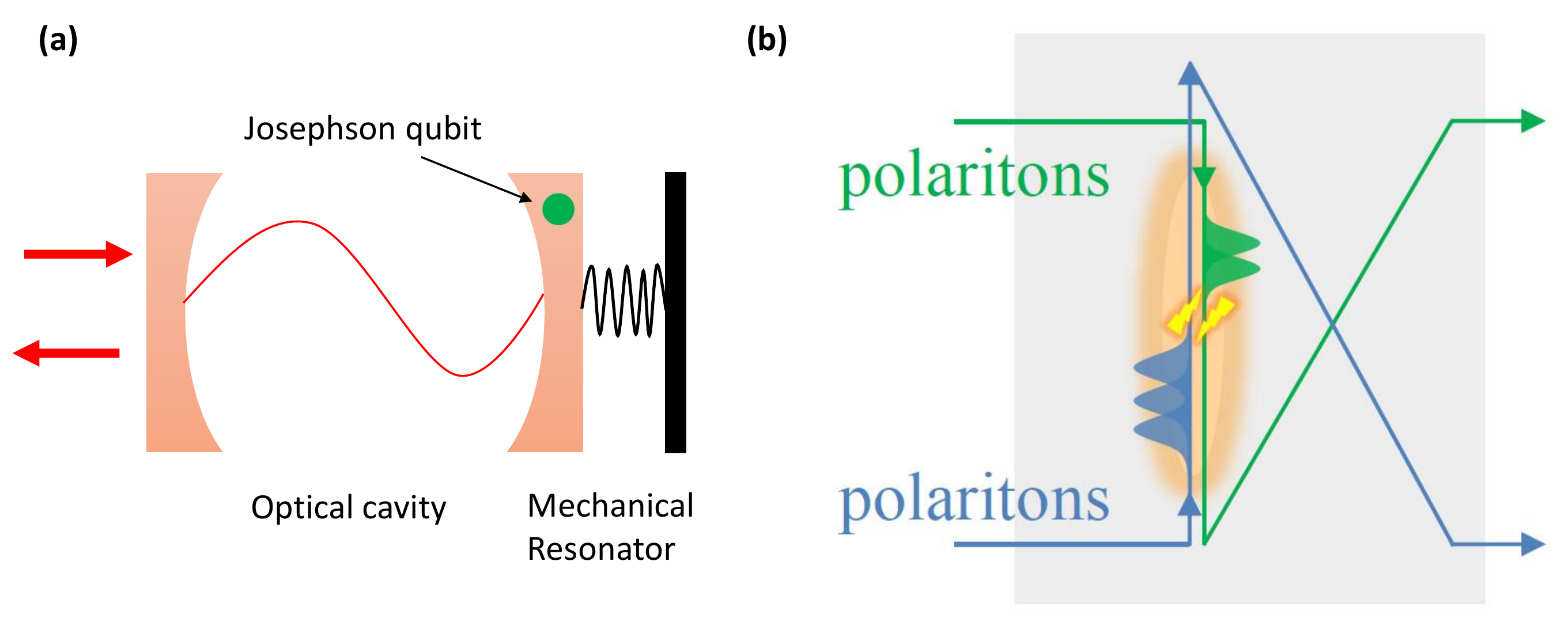,width=0.75\linewidth}
}
\caption{\label{f-cascade}
\textbf{(a) Optomechanical implementation:} Optical cavity and mechanical resonator Kerr cross-coupled by a Josephson qubit \cite{Hekkila2014PRL}. \textbf{(b) Quantum nonlinear optical implementation:} Photons entering from the two input modes counter-propagate in the form of Rydberg slow-light polaritons. Giant dipole-dipole interaction between Rydberg atoms leads to a cross-Kerr phase shift of ${n}_1{n}_2 \pi/2 $ for counter-propagating ${n}_1$ and ${n}_2$ photons. 
}
\end{figure*}

\section{Quantum distribution and work production by Cascading}\label{Cascade}

One can concentrate the energy to higher values than in  Fig. \ref{f-spheres}b by cascading the 4-mode interferometric block described above (see inset in Fig. \ref{f-spheres}c). Such a cascade can be viewed as the \textit{spatial analog of  consecutive temporal cycles of a heat engine}. In each block, the relative variance $\Delta n/\bar{n}$ is smaller than in the preceding one. As shown in Fig. \ref{f-spheres}c, this cascading yields as increasingly \textit{non-monotonic quantum number distribution} $\{p_n\} (n=0,1,2..)$ i.e. a growingly \textit{non-passive state} with \textit{net ergotropy} in the  designated working mode,  capable of delivering work \cite{Allahverdyan2004EPL,Pusz_1978,Lenard1978} (see SI).

The interferometer parameters $\chi$ and $s$ have been optimized to maximize  $\bar n-\Delta n$ (i.e., the \textit{non-passivity}) and not just $\bar n$. Although analytical formulae are tractable only for the first two moments of the quantum number distribution, we can put a bound on the \textit{non-monotonicity} of the distribution. To this end, consistently with the Jaynes principle \cite{PhysRev.106.620}, we choose the quanta number distribution of the \textit{highest entropy} that corresponds to the values of the first  two  moments. 

Invariably, the sum of the {\it single-mode entropies} of the cascaded HENLI increases (but the total entropy remains constant). The input energy fraction converted to work is below the Carnot bound. The ability to attain this bound by cascading is yet to be studied, and so is the steady-state of such a cascade.




\section{ENVISAGED EXPERIMENTAL IMPLEMENTATIONS}\label{ExperimentalProposals}
%

(a) \textit{Optomechanical setup:} The proposed scheme of HENLI can be implemented in an optomechanical setup where a microwave cavity is coupled to a mechanical resonator (Fig. \ref{f-cascade}a) with strength $g$ \cite{RevModPhys.86.1391}. The cross-Kerr coupling between the cavity and resonator can be induced by a Josephson junction qubit \cite{Hekkila2014PRL}. Due to this coupling, the cavity and mechanical resonator will have an effective cross-Kerr interaction Hamiltonian $H_{cK}=\chi a^\dagger a b^\dagger b$. As a result, the phase shift in the cavity field will depend on the number of phonons in the mechanical resonator. The estimated cross-Kerr strength for this optomechanical setup has been found to be very large, $\chi=0.25 g$ \cite{Hekkila2014PRL}.

\textit{(b) Slow-light Rydberg polaritons.:} Another feasible implementation  may invoke a cross-Kerr nonlinear element using cold atomic vapor cells. The concept of such an element stems from our theoretical ideas \cite{Shahmoon2011,Friedler} and the experiments  which demonstrated \cite{Firstenberg2013,Thompson2017,Tiarks2019} \textit{giant cross-Kerr nonlinearity} between  few-photon beams converted into Rydberg polaritons (Fig. \ref{f-cascade}b). Their dipolar interaction (either dipole blockade \cite{Thompson2017} or resonant dipolar exchange \cite{Tiarks2019}) can be tuned to yield a phase shift of $\chi=\pi/2$ between each pair of counter-propagating polaritons. 

\section{Conclusions} \label{Conclusion}
Nonlinear interferometric networks have been proposed here for the first time as \textit{fully coherent} heat engines (heat engines via nonlinear interferometry$-$HENLI). Conceptually, they allow  us to {\it treat baths as dynamical systems}, in contrast to existing classical and quantum heat engine, for which the working-medium-bath exchange has {\it never been described as a coherent process}. Such a description is indeed unfeasible for infinite/macroscopic baths, but here we consider only {\it few-mode ``baths''}, distancing ourselves from the ruling paradigm of thermodynamics. Notwithstanding its coherent-nonlinear nature, HENLI adheres to the second law and acts as a genuine heat machine cycle, albeit only few modes are involved in the operation.   

The minimal version of HENLI$-$4-mode cross-Kerr coupled network has been analyzed here to illustrate the operation principles. These nonlinearly-coupled modes replace both the working medium and the piston (working mode) so that HENLI is conceptually much simpler than a traditional heat engine. In principle, the analysis has revealed these general insights: (i) Mutual information of two-mode intensity-phase difference $(n_--\phi)$ autonomously builds up by their sampling via a linear Gaussian operation and is then reduced via nonlinear correlation that narrows down the distribution $\tilde{p}(\phi)$. The final  $\tilde{p}(\phi)$ is part of the feedforward that controls the interference of output modes and thereby steers energy to the designated mode. (ii) The two-mode Stokes operators reveal quantum nonlinear 4-mode correlations that concentrate the population in a designated mode, and render its distribution non-thermal, but also incur performance-degrading due to admixing of vacuum noise through empty ports. (iii) Appropriate cascading of 4-mode HENLI blocks can progressively augment the energy steering to and the non-passivity of the designated mode, and its non-passivity analogously to consecutive cycles.

More sophisticated versions (to be reported) involve more modes (up to 8 modes) in each block, in order to obtain MI not only on ($n_--\phi$) distribution but also on other pairs of conjugate variables, thus rendering their sampling and feedforward more complete. Such a multimode block may realize functionalities other than HE (e.g. refrigerator \cite{PhysRevLett.109.090601} or heat transistor \cite{Karimi_2017,Tahir, PhysRevE.99.042121}).

On the applied side (a) HENLI devices may give rise to new technologies of steering ambient heat (few-quanta input) in multimode networks and its conversion to quasi-coherent work output. Their practical value is their ability to {\it interfere and thereby  concentrate  energy  from independent heat channels}. This feat is impossible in conventional  heat devices, whose heat channels do not interfere. (b) The cascading process depicted here (Fig. \ref{f-spheres}c) may pave the way to \textit{manipulating and enhancing the information hidden in noisy input via controllable nonlinear operations.} This perspective is based on the remarkable fact that HENLI bears analogy to a quantum computer with continuous variables \cite{PhysRevLett.82.1784}, if inter-mode quantum correlations are accounted for, or to a semiclassical optical computer if they are neglected. (c) The feasibility of HENLI for few-photon or few-phonon input may add impetus to the creation of quantum nonlinear interference devices \cite{Shahmoon2011,Friedler,Firstenberg2013,Thompson2017,Tiarks2019,PhysRevLett.82.1784}.

By breaking away from the thermodynamic paradigm of dissipative HE operation, our long-term goal is to \textit{trace the transition from quantum or classical coherent dynamics to thermodynamics} as a function of the number of bath modes and their nonlinear coupling. Such a transition may provide the conceptual basis for a \textit{unified thermodynamic coherent approach} whereby only the input channels are thermal while the rest of the process is classically  or quantum-mechanically coherent. \\\\

\noindent
\textbf{Acknowledgments}
T.O. and \v{S}.B. are supported by the Czech Science Foundation, Grant No. 20-
27994S. \v{S}.B. is supported by project IGA-PrF-2022-005.  G.K. is supported by ISF, DFG (FOR 2724) and QUANTERA (PACE-IN). E.P., O.F. and G.K. are supported by NSF-BSF.\\

\noindent
\textbf{Author contributions}:  All  the authors contributed to the formulation of the problem, its analysis and the discussion of its results. TO, AGK, AM and NM have carried out the analytical calculations, whereas SB has performed the numerical work. EP, OF, TO and NM have analyzed the possible experimental implementations. GK, TO and AGK  have done most of the writing.\\

\noindent
\textbf{Competing interests}: The authors declare no competing interests.

\bibliography{CoherentHM}

\providecommand{\noopsort}[1]{}\providecommand{\singleletter}[1]{#1}%
\begin{thebibliography}{78}%
\makeatletter
\providecommand \@ifxundefined [1]{%
 \@ifx{#1\undefined}
}%
\providecommand \@ifnum [1]{%
 \ifnum #1\expandafter \@firstoftwo
 \else \expandafter \@secondoftwo
 \fi
}%
\providecommand \@ifx [1]{%
 \ifx #1\expandafter \@firstoftwo
 \else \expandafter \@secondoftwo
 \fi
}%
\providecommand \natexlab [1]{#1}%
\providecommand \enquote  [1]{``#1''}%
\providecommand \bibnamefont  [1]{#1}%
\providecommand \bibfnamefont [1]{#1}%
\providecommand \citenamefont [1]{#1}%
\providecommand \href@noop [0]{\@secondoftwo}%
\providecommand \href [0]{\begingroup \@sanitize@url \@href}%
\providecommand \@href[1]{\@@startlink{#1}\@@href}%
\providecommand \@@href[1]{\endgroup#1\@@endlink}%
\providecommand \@sanitize@url [0]{\catcode `\\12\catcode `\$12\catcode
  `\&12\catcode `\#12\catcode `\^12\catcode `\_12\catcode `\%12\relax}%
\providecommand \@@startlink[1]{}%
\providecommand \@@endlink[0]{}%
\providecommand \url  [0]{\begingroup\@sanitize@url \@url }%
\providecommand \@url [1]{\endgroup\@href {#1}{\urlprefix }}%
\providecommand \urlprefix  [0]{URL }%
\providecommand \Eprint [0]{\href }%
\providecommand \doibase [0]{https://doi.org/}%
\providecommand \selectlanguage [0]{\@gobble}%
\providecommand \bibinfo  [0]{\@secondoftwo}%
\providecommand \bibfield  [0]{\@secondoftwo}%
\providecommand \translation [1]{[#1]}%
\providecommand \BibitemOpen [0]{}%
\providecommand \bibitemStop [0]{}%
\providecommand \bibitemNoStop [0]{.\EOS\space}%
\providecommand \EOS [0]{\spacefactor3000\relax}%
\providecommand \BibitemShut  [1]{\csname bibitem#1\endcsname}%
\let\auto@bib@innerbib\@empty
\bibitem [{\citenamefont {Callen}(1985)}]{callenbook}%
  \BibitemOpen
  \bibfield  {author} {\bibinfo {author} {\bibfnamefont {H.~B.}\ \bibnamefont
  {Callen}},\ }\href@noop {} {\emph {\bibinfo {title} {{Thermodynamics and an
  Introduction to Thermostatistics}}}}\ (\bibinfo  {publisher} {John Wiley \&
  Sons, Inc.},\ \bibinfo {year} {1985})\BibitemShut {NoStop}%
\bibitem [{\citenamefont {Kondepudi}\ and\ \citenamefont
  {Prigogine}(2014)}]{KondepudiPrigogine}%
  \BibitemOpen
  \bibfield  {author} {\bibinfo {author} {\bibfnamefont {D.}~\bibnamefont
  {Kondepudi}}\ and\ \bibinfo {author} {\bibfnamefont {I.}~\bibnamefont
  {Prigogine}},\ }\href@noop {} {\emph {\bibinfo {title} {Modern
  Thermodynamics: From Heat Engines to Dissipative Structures, 2nd edition}}}\
  (\bibinfo  {publisher} {Wiley},\ \bibinfo {address} {Chichester},\ \bibinfo
  {year} {2014})\BibitemShut {NoStop}%
\bibitem [{\citenamefont {Kosloff}(2013)}]{Kosloff_2013}%
  \BibitemOpen
  \bibfield  {author} {\bibinfo {author} {\bibfnamefont {R.}~\bibnamefont
  {Kosloff}},\ }\bibfield  {title} {\bibinfo {title} {Quantum thermodynamics: A
  dynamical viewpoint},\ }\href {https://doi.org/10.3390/e15062100} {\bibfield
  {journal} {\bibinfo  {journal} {Entropy}\ }\textbf {\bibinfo {volume} {15}},\
  \bibinfo {pages} {2100} (\bibinfo {year} {2013})}\BibitemShut {NoStop}%
\bibitem [{\citenamefont {Binder}\ \emph {et~al.}(2018)\citenamefont {Binder},
  \citenamefont {Correa}, \citenamefont {Gogolin}, \citenamefont {Anders},\
  and\ \citenamefont {Adesso}}]{BinderBook}%
  \BibitemOpen
  \bibfield  {author} {\bibinfo {author} {\bibfnamefont {F.}~\bibnamefont
  {Binder}}, \bibinfo {author} {\bibfnamefont {L.}~\bibnamefont {Correa}},
  \bibinfo {author} {\bibfnamefont {C.}~\bibnamefont {Gogolin}}, \bibinfo
  {author} {\bibfnamefont {J.}~\bibnamefont {Anders}},\ and\ \bibinfo {author}
  {\bibfnamefont {G.~e.}\ \bibnamefont {Adesso}},\ }\href@noop {} {\emph
  {\bibinfo {title} {Thermodynamics in the Quantum Regime: Fundamental Aspects
  and New Directions}}}\ (\bibinfo  {publisher} {Springer},\ \bibinfo {year}
  {2018})\BibitemShut {NoStop}%
\bibitem [{\citenamefont {Gelbwaser-Klimovsky}\ \emph
  {et~al.}(2015)\citenamefont {Gelbwaser-Klimovsky}, \citenamefont {Niedenzu},\
  and\ \citenamefont {Kurizki}}]{David2015AdvAtMolOpt}%
  \BibitemOpen
  \bibfield  {author} {\bibinfo {author} {\bibfnamefont {D.}~\bibnamefont
  {Gelbwaser-Klimovsky}}, \bibinfo {author} {\bibfnamefont {W.}~\bibnamefont
  {Niedenzu}},\ and\ \bibinfo {author} {\bibfnamefont {G.}~\bibnamefont
  {Kurizki}},\ }\bibfield  {title} {\bibinfo {title} {Chapter twelve -
  thermodynamics of quantum systems under dynamical control}\ }(\bibinfo
  {publisher} {Academic Press},\ \bibinfo {year} {2015})\ pp.\ \bibinfo {pages}
  {329--407}\BibitemShut {NoStop}%
\bibitem [{\citenamefont {Kurizki}\ and\ \citenamefont
  {Kofman}(2022)}]{Kurizkibook}%
  \BibitemOpen
  \bibfield  {author} {\bibinfo {author} {\bibfnamefont {G.}~\bibnamefont
  {Kurizki}}\ and\ \bibinfo {author} {\bibfnamefont {A.~G.}\ \bibnamefont
  {Kofman}},\ }\href@noop {} {\emph {\bibinfo {title} {Thermodynamics and
  Control of Open Quantum Systems}}}\ (\bibinfo  {publisher} {Cambridge
  University Press},\ \bibinfo {year} {2022})\BibitemShut {NoStop}%
\bibitem [{\citenamefont {Scovil}\ and\ \citenamefont
  {Schulz-DuBois}(1959)}]{PhysRevLett.2.262}%
  \BibitemOpen
  \bibfield  {author} {\bibinfo {author} {\bibfnamefont {H.~E.~D.}\
  \bibnamefont {Scovil}}\ and\ \bibinfo {author} {\bibfnamefont {E.~O.}\
  \bibnamefont {Schulz-DuBois}},\ }\bibfield  {title} {\bibinfo {title}
  {Three-level masers as heat engines},\ }\href
  {https://doi.org/10.1103/PhysRevLett.2.262} {\bibfield  {journal} {\bibinfo
  {journal} {Phys. Rev. Lett.}\ }\textbf {\bibinfo {volume} {2}},\ \bibinfo
  {pages} {262} (\bibinfo {year} {1959})}\BibitemShut {NoStop}%
\bibitem [{\citenamefont {Alicki}(1979)}]{Alicki_1979}%
  \BibitemOpen
  \bibfield  {author} {\bibinfo {author} {\bibfnamefont {R.}~\bibnamefont
  {Alicki}},\ }\bibfield  {title} {\bibinfo {title} {The quantum open system as
  a model of the heat engine},\ }\href
  {https://doi.org/10.1088/0305-4470/12/5/007} {\bibfield  {journal} {\bibinfo
  {journal} {Journal of Physics A: Mathematical and General}\ }\textbf
  {\bibinfo {volume} {12}},\ \bibinfo {pages} {L103} (\bibinfo {year}
  {1979})}\BibitemShut {NoStop}%
\bibitem [{\citenamefont {Huang}\ \emph {et~al.}(2012)\citenamefont {Huang},
  \citenamefont {Wang},\ and\ \citenamefont {Yi}}]{PhysRevE.86.051105}%
  \BibitemOpen
  \bibfield  {author} {\bibinfo {author} {\bibfnamefont {X.~L.}\ \bibnamefont
  {Huang}}, \bibinfo {author} {\bibfnamefont {T.}~\bibnamefont {Wang}},\ and\
  \bibinfo {author} {\bibfnamefont {X.~X.}\ \bibnamefont {Yi}},\ }\bibfield
  {title} {\bibinfo {title} {Effects of reservoir squeezing on quantum systems
  and work extraction},\ }\href {https://doi.org/10.1103/PhysRevE.86.051105}
  {\bibfield  {journal} {\bibinfo  {journal} {Phys. Rev. E}\ }\textbf {\bibinfo
  {volume} {86}},\ \bibinfo {pages} {051105} (\bibinfo {year}
  {2012})}\BibitemShut {NoStop}%
\bibitem [{\citenamefont {Scully}\ \emph {et~al.}(2003)\citenamefont {Scully},
  \citenamefont {Zubairy}, \citenamefont {Agarwal},\ and\ \citenamefont
  {Walther}}]{Scully862}%
  \BibitemOpen
  \bibfield  {author} {\bibinfo {author} {\bibfnamefont {M.~O.}\ \bibnamefont
  {Scully}}, \bibinfo {author} {\bibfnamefont {M.~S.}\ \bibnamefont {Zubairy}},
  \bibinfo {author} {\bibfnamefont {G.~S.}\ \bibnamefont {Agarwal}},\ and\
  \bibinfo {author} {\bibfnamefont {H.}~\bibnamefont {Walther}},\ }\bibfield
  {title} {\bibinfo {title} {Extracting work from a single heat bath via
  vanishing quantum coherence},\ }\href
  {https://doi.org/10.1126/science.1078955} {\bibfield  {journal} {\bibinfo
  {journal} {Science}\ }\textbf {\bibinfo {volume} {299}},\ \bibinfo {pages}
  {862} (\bibinfo {year} {2003})}\BibitemShut {NoStop}%
\bibitem [{\citenamefont {Uzdin}\ \emph {et~al.}(2015)\citenamefont {Uzdin},
  \citenamefont {Levy},\ and\ \citenamefont {Kosloff}}]{PhysRevX.5.031044}%
  \BibitemOpen
  \bibfield  {author} {\bibinfo {author} {\bibfnamefont {R.}~\bibnamefont
  {Uzdin}}, \bibinfo {author} {\bibfnamefont {A.}~\bibnamefont {Levy}},\ and\
  \bibinfo {author} {\bibfnamefont {R.}~\bibnamefont {Kosloff}},\ }\bibfield
  {title} {\bibinfo {title} {Equivalence of quantum heat machines, and
  quantum-thermodynamic signatures},\ }\href
  {https://doi.org/10.1103/PhysRevX.5.031044} {\bibfield  {journal} {\bibinfo
  {journal} {Phys. Rev. X}\ }\textbf {\bibinfo {volume} {5}},\ \bibinfo {pages}
  {031044} (\bibinfo {year} {2015})}\BibitemShut {NoStop}%
\bibitem [{\citenamefont {Campisi}\ and\ \citenamefont
  {Fazio}(2016)}]{Campisi}%
  \BibitemOpen
  \bibfield  {author} {\bibinfo {author} {\bibfnamefont {M.}~\bibnamefont
  {Campisi}}\ and\ \bibinfo {author} {\bibfnamefont {R.}~\bibnamefont
  {Fazio}},\ }\bibfield  {title} {\bibinfo {title} {The power of a critical
  heat engine},\ }\href {https://doi.org/10.1038/ncomms11895} {\bibfield
  {journal} {\bibinfo  {journal} {Nature Communications}\ }\textbf {\bibinfo
  {volume} {7}},\ \bibinfo {pages} {11895} (\bibinfo {year}
  {2016})}\BibitemShut {NoStop}%
\bibitem [{\citenamefont {{Ro{\ss}nagel}}\ \emph {et~al.}(2016)\citenamefont
  {{Ro{\ss}nagel}}, \citenamefont {{Dawkins}}, \citenamefont {{Tolazzi}},
  \citenamefont {{Abah}}, \citenamefont {{Lutz}}, \citenamefont
  {{Schmidt-Kaler}},\ and\ \citenamefont {{Singer}}}]{FTEE}%
  \BibitemOpen
  \bibfield  {author} {\bibinfo {author} {\bibfnamefont {J.}~\bibnamefont
  {{Ro{\ss}nagel}}}, \bibinfo {author} {\bibfnamefont {S.~T.}\ \bibnamefont
  {{Dawkins}}}, \bibinfo {author} {\bibfnamefont {K.~N.}\ \bibnamefont
  {{Tolazzi}}}, \bibinfo {author} {\bibfnamefont {O.}~\bibnamefont {{Abah}}},
  \bibinfo {author} {\bibfnamefont {E.}~\bibnamefont {{Lutz}}}, \bibinfo
  {author} {\bibfnamefont {F.}~\bibnamefont {{Schmidt-Kaler}}},\ and\ \bibinfo
  {author} {\bibfnamefont {K.}~\bibnamefont {{Singer}}},\ }\bibfield  {title}
  {\bibinfo {title} {{A single-atom heat engine}},\ }\href
  {https://doi.org/10.1126/science.aad6320} {\bibfield  {journal} {\bibinfo
  {journal} {Science}\ }\textbf {\bibinfo {volume} {352}},\ \bibinfo {pages}
  {325} (\bibinfo {year} {2016})},\ \Eprint {https://arxiv.org/abs/1510.03681}
  {arXiv:1510.03681 [cond-mat.stat-mech]} \BibitemShut {NoStop}%
\bibitem [{\citenamefont {Gelbwaser-Klimovsky}\ \emph
  {et~al.}(2013{\natexlab{a}})\citenamefont {Gelbwaser-Klimovsky},
  \citenamefont {Alicki},\ and\ \citenamefont {Kurizki}}]{Gelbwaser_2013_b}%
  \BibitemOpen
  \bibfield  {author} {\bibinfo {author} {\bibfnamefont {D.}~\bibnamefont
  {Gelbwaser-Klimovsky}}, \bibinfo {author} {\bibfnamefont {R.}~\bibnamefont
  {Alicki}},\ and\ \bibinfo {author} {\bibfnamefont {G.}~\bibnamefont
  {Kurizki}},\ }\bibfield  {title} {\bibinfo {title} {Work and energy gain of
  heat-pumped quantized amplifiers},\ }\href
  {http://stacks.iop.org/0295-5075/103/i=6/a=60005} {\bibfield  {journal}
  {\bibinfo  {journal} {EPL}\ }\textbf {\bibinfo {volume} {103}},\ \bibinfo
  {pages} {60005} (\bibinfo {year} {2013)}{\natexlab{a}})}\BibitemShut
  {NoStop}%
\bibitem [{\citenamefont {Gelbwaser-Klimovsky}\ and\ \citenamefont
  {Kurizki}(2014)}]{PRE2014}%
  \BibitemOpen
  \bibfield  {author} {\bibinfo {author} {\bibfnamefont {D.}~\bibnamefont
  {Gelbwaser-Klimovsky}}\ and\ \bibinfo {author} {\bibfnamefont
  {G.}~\bibnamefont {Kurizki}},\ }\bibfield  {title} {\bibinfo {title}
  {Heat-machine control by quantum-state preparation: From quantum engines to
  refrigerators},\ }\href {https://doi.org/10.1103/PhysRevE.90.022102}
  {\bibfield  {journal} {\bibinfo  {journal} {Phys. Rev. E}\ }\textbf {\bibinfo
  {volume} {90}},\ \bibinfo {pages} {022102} (\bibinfo {year}
  {2014})}\BibitemShut {NoStop}%
\bibitem [{\citenamefont {Niedenzu}\ \emph {et~al.}(2016)\citenamefont
  {Niedenzu}, \citenamefont {Gelbwaser-Klimovsky}, \citenamefont {Kofman},\
  and\ \citenamefont {Kurizki}}]{Niedenzu_2016}%
  \BibitemOpen
  \bibfield  {author} {\bibinfo {author} {\bibfnamefont {W.}~\bibnamefont
  {Niedenzu}}, \bibinfo {author} {\bibfnamefont {D.}~\bibnamefont
  {Gelbwaser-Klimovsky}}, \bibinfo {author} {\bibfnamefont {A.~G.}\
  \bibnamefont {Kofman}},\ and\ \bibinfo {author} {\bibfnamefont
  {G.}~\bibnamefont {Kurizki}},\ }\bibfield  {title} {\bibinfo {title} {On the
  operation of machines powered by quantum non-thermal baths},\ }\href
  {https://doi.org/10.1088/1367-2630/18/8/083012} {\bibfield  {journal}
  {\bibinfo  {journal} {New Journal of Physics}\ }\textbf {\bibinfo {volume}
  {18}},\ \bibinfo {pages} {083012} (\bibinfo {year} {2016})}\BibitemShut
  {NoStop}%
\bibitem [{\citenamefont {Da\u{g}}\ \emph {et~al.}(2016)\citenamefont
  {Da\u{g}}, \citenamefont {Niedenzu}, \citenamefont
  {M\"{u}stecapl{\i}o\u{g}lu},\ and\ \citenamefont {Kurizki}}]{e18070244}%
  \BibitemOpen
  \bibfield  {author} {\bibinfo {author} {\bibfnamefont {C.~B.}\ \bibnamefont
  {Da\u{g}}}, \bibinfo {author} {\bibfnamefont {W.}~\bibnamefont {Niedenzu}},
  \bibinfo {author} {\bibfnamefont {O.~E.}\ \bibnamefont
  {M\"{u}stecapl{\i}o\u{g}lu}},\ and\ \bibinfo {author} {\bibfnamefont
  {G.}~\bibnamefont {Kurizki}},\ }\bibfield  {title} {\bibinfo {title}
  {Multiatom quantum coherences in micromasers as fuel for thermal and
  nonthermal machines},\ }\bibfield  {journal} {\bibinfo  {journal} {Entropy}\
  }\textbf {\bibinfo {volume} {18}},\ \href {https://doi.org/10.3390/e18070244}
  {10.3390/e18070244} (\bibinfo {year} {2016})\BibitemShut {NoStop}%
\bibitem [{\citenamefont {Mukherjee}\ \emph {et~al.}(2016)\citenamefont
  {Mukherjee}, \citenamefont {Niedenzu}, \citenamefont {Kofman},\ and\
  \citenamefont {Kurizki}}]{PhysRevE.94.062109}%
  \BibitemOpen
  \bibfield  {author} {\bibinfo {author} {\bibfnamefont {V.}~\bibnamefont
  {Mukherjee}}, \bibinfo {author} {\bibfnamefont {W.}~\bibnamefont {Niedenzu}},
  \bibinfo {author} {\bibfnamefont {A.~G.}\ \bibnamefont {Kofman}},\ and\
  \bibinfo {author} {\bibfnamefont {G.}~\bibnamefont {Kurizki}},\ }\bibfield
  {title} {\bibinfo {title} {Speed and efficiency limits of multilevel
  incoherent heat engines},\ }\href
  {https://doi.org/10.1103/PhysRevE.94.062109} {\bibfield  {journal} {\bibinfo
  {journal} {Phys. Rev. E}\ }\textbf {\bibinfo {volume} {94}},\ \bibinfo
  {pages} {062109} (\bibinfo {year} {2016})}\BibitemShut {NoStop}%
\bibitem [{\citenamefont {Ghosh}\ \emph {et~al.}(2017)\citenamefont {Ghosh},
  \citenamefont {Latune}, \citenamefont {Davidovich},\ and\ \citenamefont
  {Kurizki}}]{Ghosh2017}%
  \BibitemOpen
  \bibfield  {author} {\bibinfo {author} {\bibfnamefont {A.}~\bibnamefont
  {Ghosh}}, \bibinfo {author} {\bibfnamefont {C.~L.}\ \bibnamefont {Latune}},
  \bibinfo {author} {\bibfnamefont {L.}~\bibnamefont {Davidovich}},\ and\
  \bibinfo {author} {\bibfnamefont {G.}~\bibnamefont {Kurizki}},\ }\bibfield
  {title} {\bibinfo {title} {Catalysis of heat-to-work conversion in quantum
  machines},\ }\href {https://doi.org/10.1073/pnas.1711381114} {\bibfield
  {journal} {\bibinfo  {journal} {Proceedings of the National Academy of
  Sciences}\ }\textbf {\bibinfo {volume} {114}},\ \bibinfo {pages} {12156}
  (\bibinfo {year} {2017})}\BibitemShut {NoStop}%
\bibitem [{\citenamefont {Ghosh}\ \emph {et~al.}(2018)\citenamefont {Ghosh},
  \citenamefont {Gelbwaser-Klimovsky}, \citenamefont {Niedenzu}, \citenamefont
  {Lvovsky}, \citenamefont {Mazets}, \citenamefont {Scully},\ and\
  \citenamefont {Kurizki}}]{Ghosh2018}%
  \BibitemOpen
  \bibfield  {author} {\bibinfo {author} {\bibfnamefont {A.}~\bibnamefont
  {Ghosh}}, \bibinfo {author} {\bibfnamefont {D.}~\bibnamefont
  {Gelbwaser-Klimovsky}}, \bibinfo {author} {\bibfnamefont {W.}~\bibnamefont
  {Niedenzu}}, \bibinfo {author} {\bibfnamefont {A.~I.}\ \bibnamefont
  {Lvovsky}}, \bibinfo {author} {\bibfnamefont {I.}~\bibnamefont {Mazets}},
  \bibinfo {author} {\bibfnamefont {M.~O.}\ \bibnamefont {Scully}},\ and\
  \bibinfo {author} {\bibfnamefont {G.}~\bibnamefont {Kurizki}},\ }\bibfield
  {title} {\bibinfo {title} {Two-level masers as heat-to-work converters},\
  }\href {https://doi.org/10.1073/pnas.1805354115} {\bibfield  {journal}
  {\bibinfo  {journal} {Proceedings of the National Academy of Sciences}\
  }\textbf {\bibinfo {volume} {115}},\ \bibinfo {pages} {9941} (\bibinfo {year}
  {2018})}\BibitemShut {NoStop}%
\bibitem [{\citenamefont {Niedenzu}\ \emph {et~al.}(2018)\citenamefont
  {Niedenzu}, \citenamefont {Mukherjee}, \citenamefont {Ghosh}, \citenamefont
  {Kofman},\ and\ \citenamefont {Kurizki}}]{niedenzu18quantum}%
  \BibitemOpen
  \bibfield  {author} {\bibinfo {author} {\bibfnamefont {W.}~\bibnamefont
  {Niedenzu}}, \bibinfo {author} {\bibfnamefont {V.}~\bibnamefont {Mukherjee}},
  \bibinfo {author} {\bibfnamefont {A.}~\bibnamefont {Ghosh}}, \bibinfo
  {author} {\bibfnamefont {A.~G.}\ \bibnamefont {Kofman}},\ and\ \bibinfo
  {author} {\bibfnamefont {G.}~\bibnamefont {Kurizki}},\ }\bibfield  {title}
  {\bibinfo {title} {Quantum engine efficiency bound beyond the second law of
  thermodynamics},\ }\href {https://doi.org/10.1038/s41467-017-01991-6}
  {\bibfield  {journal} {\bibinfo  {journal} {Nature Communications}\ }\textbf
  {\bibinfo {volume} {9}},\ \bibinfo {pages} {165} (\bibinfo {year}
  {2018})}\BibitemShut {NoStop}%
\bibitem [{\citenamefont {Dillenschneider}\ and\ \citenamefont
  {Lutz}(2009)}]{Dillenschneider_2009}%
  \BibitemOpen
  \bibfield  {author} {\bibinfo {author} {\bibfnamefont {R.}~\bibnamefont
  {Dillenschneider}}\ and\ \bibinfo {author} {\bibfnamefont {E.}~\bibnamefont
  {Lutz}},\ }\bibfield  {title} {\bibinfo {title} {Energetics of quantum
  correlations},\ }\href {https://doi.org/10.1209/0295-5075/88/50003}
  {\bibfield  {journal} {\bibinfo  {journal} {{EPL} (Europhysics Letters)}\
  }\textbf {\bibinfo {volume} {88}},\ \bibinfo {pages} {50003} (\bibinfo {year}
  {2009})}\BibitemShut {NoStop}%
\bibitem [{\citenamefont {Abah}\ and\ \citenamefont {Lutz}(2014)}]{Abah_2014}%
  \BibitemOpen
  \bibfield  {author} {\bibinfo {author} {\bibfnamefont {O.}~\bibnamefont
  {Abah}}\ and\ \bibinfo {author} {\bibfnamefont {E.}~\bibnamefont {Lutz}},\
  }\bibfield  {title} {\bibinfo {title} {Efficiency of heat engines coupled to
  nonequilibrium reservoirs},\ }\href
  {https://doi.org/10.1209/0295-5075/106/20001} {\bibfield  {journal} {\bibinfo
   {journal} {{EPL} (Europhysics Letters)}\ }\textbf {\bibinfo {volume}
  {106}},\ \bibinfo {pages} {20001} (\bibinfo {year} {2014})}\BibitemShut
  {NoStop}%
\bibitem [{\citenamefont {Ro\ss{}nagel}\ \emph {et~al.}(2014)\citenamefont
  {Ro\ss{}nagel}, \citenamefont {Abah}, \citenamefont {Schmidt-Kaler},
  \citenamefont {Singer},\ and\ \citenamefont {Lutz}}]{FTE4}%
  \BibitemOpen
  \bibfield  {author} {\bibinfo {author} {\bibfnamefont {J.}~\bibnamefont
  {Ro\ss{}nagel}}, \bibinfo {author} {\bibfnamefont {O.}~\bibnamefont {Abah}},
  \bibinfo {author} {\bibfnamefont {F.}~\bibnamefont {Schmidt-Kaler}}, \bibinfo
  {author} {\bibfnamefont {K.}~\bibnamefont {Singer}},\ and\ \bibinfo {author}
  {\bibfnamefont {E.}~\bibnamefont {Lutz}},\ }\bibfield  {title} {\bibinfo
  {title} {Nanoscale heat engine beyond the carnot limit},\ }\href
  {https://doi.org/10.1103/PhysRevLett.112.030602} {\bibfield  {journal}
  {\bibinfo  {journal} {Phys. Rev. Lett.}\ }\textbf {\bibinfo {volume} {112}},\
  \bibinfo {pages} {030602} (\bibinfo {year} {2014})}\BibitemShut {NoStop}%
\bibitem [{\citenamefont {Hardal}\ and\ \citenamefont
  {Müstecaplıoğlu}(2015)}]{Hardal2015ScRep}%
  \BibitemOpen
  \bibfield  {author} {\bibinfo {author} {\bibfnamefont {A.~.~C.}\ \bibnamefont
  {Hardal}}\ and\ \bibinfo {author} {\bibfnamefont {z.~E.}\ \bibnamefont
  {Müstecaplıoğlu}},\ }\bibfield  {title} {\bibinfo {title} {Superradiant
  quantum heat engine},\ }\href {https://doi.org/10.1038/srep12953} {\bibfield
  {journal} {\bibinfo  {journal} {Scientific Reports}\ }\textbf {\bibinfo
  {volume} {5}},\ \bibinfo {pages} {12953} (\bibinfo {year}
  {2015})}\BibitemShut {NoStop}%
\bibitem [{\citenamefont {Klaers}\ \emph {et~al.}(2017)\citenamefont {Klaers},
  \citenamefont {Faelt}, \citenamefont {Imamoglu},\ and\ \citenamefont
  {Togan}}]{PhysRevX.7.031044}%
  \BibitemOpen
  \bibfield  {author} {\bibinfo {author} {\bibfnamefont {J.}~\bibnamefont
  {Klaers}}, \bibinfo {author} {\bibfnamefont {S.}~\bibnamefont {Faelt}},
  \bibinfo {author} {\bibfnamefont {A.}~\bibnamefont {Imamoglu}},\ and\
  \bibinfo {author} {\bibfnamefont {E.}~\bibnamefont {Togan}},\ }\bibfield
  {title} {\bibinfo {title} {Squeezed thermal reservoirs as a resource for a
  nanomechanical engine beyond the carnot limit},\ }\href
  {https://doi.org/10.1103/PhysRevX.7.031044} {\bibfield  {journal} {\bibinfo
  {journal} {Phys. Rev. X}\ }\textbf {\bibinfo {volume} {7}},\ \bibinfo {pages}
  {031044} (\bibinfo {year} {2017})}\BibitemShut {NoStop}%
\bibitem [{\citenamefont {Mukhopadhyay}\ \emph {et~al.}(2018)\citenamefont
  {Mukhopadhyay}, \citenamefont {Misra}, \citenamefont {Bhattacharya},\ and\
  \citenamefont {Pati}}]{Mukhopadhyay2018PRE}%
  \BibitemOpen
  \bibfield  {author} {\bibinfo {author} {\bibfnamefont {C.}~\bibnamefont
  {Mukhopadhyay}}, \bibinfo {author} {\bibfnamefont {A.}~\bibnamefont {Misra}},
  \bibinfo {author} {\bibfnamefont {S.}~\bibnamefont {Bhattacharya}},\ and\
  \bibinfo {author} {\bibfnamefont {A.~K.}\ \bibnamefont {Pati}},\ }\bibfield
  {title} {\bibinfo {title} {Quantum speed limit constraints on a nanoscale
  autonomous refrigerator},\ }\href
  {https://doi.org/10.1103/PhysRevE.97.062116} {\bibfield  {journal} {\bibinfo
  {journal} {Phys. Rev. E}\ }\textbf {\bibinfo {volume} {97}},\ \bibinfo
  {pages} {062116} (\bibinfo {year} {2018})}\BibitemShut {NoStop}%
\bibitem [{\citenamefont {Misra}\ \emph {et~al.}(2015)\citenamefont {Misra},
  \citenamefont {Singh}, \citenamefont {Bera},\ and\ \citenamefont
  {Rajagopal}}]{CarnotPRE}%
  \BibitemOpen
  \bibfield  {author} {\bibinfo {author} {\bibfnamefont {A.}~\bibnamefont
  {Misra}}, \bibinfo {author} {\bibfnamefont {U.}~\bibnamefont {Singh}},
  \bibinfo {author} {\bibfnamefont {M.~N.}\ \bibnamefont {Bera}},\ and\
  \bibinfo {author} {\bibfnamefont {A.~K.}\ \bibnamefont {Rajagopal}},\
  }\bibfield  {title} {\bibinfo {title} {Quantum r\'enyi relative entropies
  affirm universality of thermodynamics},\ }\href
  {https://doi.org/10.1103/PhysRevE.92.042161} {\bibfield  {journal} {\bibinfo
  {journal} {Phys. Rev. E}\ }\textbf {\bibinfo {volume} {92}},\ \bibinfo
  {pages} {042161} (\bibinfo {year} {2015})}\BibitemShut {NoStop}%
\bibitem [{\citenamefont {Das}\ \emph {et~al.}(2019)\citenamefont {Das},
  \citenamefont {Misra}, \citenamefont {Pal}, \citenamefont {Sen(De)},\ and\
  \citenamefont {Sen}}]{AREPL}%
  \BibitemOpen
  \bibfield  {author} {\bibinfo {author} {\bibfnamefont {S.}~\bibnamefont
  {Das}}, \bibinfo {author} {\bibfnamefont {A.}~\bibnamefont {Misra}}, \bibinfo
  {author} {\bibfnamefont {A.~K.}\ \bibnamefont {Pal}}, \bibinfo {author}
  {\bibfnamefont {A.}~\bibnamefont {Sen(De)}},\ and\ \bibinfo {author}
  {\bibfnamefont {U.}~\bibnamefont {Sen}},\ }\bibfield  {title} {\bibinfo
  {title} {Necessarily transient quantum refrigerator},\ }\href
  {https://doi.org/10.1209/0295-5075/125/20007} {\bibfield  {journal} {\bibinfo
   {journal} {Euro Phys. Lett.}\ }\textbf {\bibinfo {volume} {125}},\ \bibinfo
  {pages} {20007} (\bibinfo {year} {2019})}\BibitemShut {NoStop}%
\bibitem [{\citenamefont {Mitchison}\ \emph {et~al.}(2015)\citenamefont
  {Mitchison}, \citenamefont {Woods}, \citenamefont {Prior},\ and\
  \citenamefont {Huber}}]{Huber}%
  \BibitemOpen
  \bibfield  {author} {\bibinfo {author} {\bibfnamefont {M.~T.}\ \bibnamefont
  {Mitchison}}, \bibinfo {author} {\bibfnamefont {M.~P.}\ \bibnamefont
  {Woods}}, \bibinfo {author} {\bibfnamefont {J.}~\bibnamefont {Prior}},\ and\
  \bibinfo {author} {\bibfnamefont {M.}~\bibnamefont {Huber}},\ }\bibfield
  {title} {\bibinfo {title} {Coherence-assisted single-shot cooling by quantum
  absorption refrigerators},\ }\href
  {https://doi.org/10.1088/1367-2630/17/11/115013} {\bibfield  {journal}
  {\bibinfo  {journal} {New Journal of Physics}\ }\textbf {\bibinfo {volume}
  {17}},\ \bibinfo {pages} {115013} (\bibinfo {year} {2015})}\BibitemShut
  {NoStop}%
\bibitem [{\citenamefont {Gardas}\ and\ \citenamefont
  {Deffner}(2015)}]{Deffner}%
  \BibitemOpen
  \bibfield  {author} {\bibinfo {author} {\bibfnamefont {B.}~\bibnamefont
  {Gardas}}\ and\ \bibinfo {author} {\bibfnamefont {S.}~\bibnamefont
  {Deffner}},\ }\bibfield  {title} {\bibinfo {title} {Thermodynamic
  universality of quantum carnot engines},\ }\href
  {https://doi.org/10.1103/PhysRevE.92.042126} {\bibfield  {journal} {\bibinfo
  {journal} {Phys. Rev. E}\ }\textbf {\bibinfo {volume} {92}},\ \bibinfo
  {pages} {042126} (\bibinfo {year} {2015})}\BibitemShut {NoStop}%
\bibitem [{\citenamefont {Allahverdyan}\ \emph {et~al.}(2004)\citenamefont
  {Allahverdyan}, \citenamefont {Balian},\ and\ \citenamefont
  {Nieuwenhuizen}}]{Allahverdyan2004EPL}%
  \BibitemOpen
  \bibfield  {author} {\bibinfo {author} {\bibfnamefont {A.~E.}\ \bibnamefont
  {Allahverdyan}}, \bibinfo {author} {\bibfnamefont {R.}~\bibnamefont
  {Balian}},\ and\ \bibinfo {author} {\bibfnamefont {T.~M.}\ \bibnamefont
  {Nieuwenhuizen}},\ }\bibfield  {title} {\bibinfo {title} {Maximal work
  extraction from finite quantum systems},\ }\href
  {https://doi.org/10.1209/epl/i2004-10101-2} {\bibfield  {journal} {\bibinfo
  {journal} {Europhysics Letters ({EPL})}\ }\textbf {\bibinfo {volume} {67}},\
  \bibinfo {pages} {565} (\bibinfo {year} {2004})}\BibitemShut {NoStop}%
\bibitem [{\citenamefont {Pusz}\ and\ \citenamefont
  {Woronowicz}(1978)}]{Pusz_1978}%
  \BibitemOpen
  \bibfield  {author} {\bibinfo {author} {\bibfnamefont {W.}~\bibnamefont
  {Pusz}}\ and\ \bibinfo {author} {\bibfnamefont {S.~L.}\ \bibnamefont
  {Woronowicz}},\ }\bibfield  {title} {\bibinfo {title} {Passive states and kms
  states for general quantum systems},\ }\href
  {http://projecteuclid.org/euclid.cmp/1103901491} {\bibfield  {journal}
  {\bibinfo  {journal} {Comm. Math. Phys.}\ }\textbf {\bibinfo {volume} {58}},\
  \bibinfo {pages} {273} (\bibinfo {year} {1978})}\BibitemShut {NoStop}%
\bibitem [{\citenamefont {Lenard}(1978)}]{Lenard1978}%
  \BibitemOpen
  \bibfield  {author} {\bibinfo {author} {\bibfnamefont {A.}~\bibnamefont
  {Lenard}},\ }\bibfield  {title} {\bibinfo {title} {Thermodynamical proof of
  the gibbs formula for elementary quantum systems},\ }\href
  {https://doi.org/10.1007/BF01011769} {\bibfield  {journal} {\bibinfo
  {journal} {Journal of Statistical Physics}\ }\textbf {\bibinfo {volume}
  {19}},\ \bibinfo {pages} {575} (\bibinfo {year} {1978})}\BibitemShut
  {NoStop}%
\bibitem [{\citenamefont {Lutz}\ and\ \citenamefont
  {Ciliberto}(2015)}]{lutz2015information}%
  \BibitemOpen
  \bibfield  {author} {\bibinfo {author} {\bibfnamefont {E.}~\bibnamefont
  {Lutz}}\ and\ \bibinfo {author} {\bibfnamefont {S.}~\bibnamefont
  {Ciliberto}},\ }\bibfield  {title} {\bibinfo {title} {Information: From
  {Maxwell’s} demon to {Landauer’s} eraser},\ }\href
  {https://doi.org/10.1063/PT.3.2912} {\bibfield  {journal} {\bibinfo
  {journal} {Phys. Today}\ }\textbf {\bibinfo {volume} {68}},\ \bibinfo {pages}
  {30} (\bibinfo {year} {2015})}\BibitemShut {NoStop}%
\bibitem [{\citenamefont {Vidrighin}\ \emph {et~al.}(2016)\citenamefont
  {Vidrighin}, \citenamefont {Dahlsten}, \citenamefont {Barbieri},
  \citenamefont {Kim}, \citenamefont {Vedral},\ and\ \citenamefont
  {Walmsley}}]{Vid2016}%
  \BibitemOpen
  \bibfield  {author} {\bibinfo {author} {\bibfnamefont {M.~D.}\ \bibnamefont
  {Vidrighin}}, \bibinfo {author} {\bibfnamefont {O.}~\bibnamefont {Dahlsten}},
  \bibinfo {author} {\bibfnamefont {M.}~\bibnamefont {Barbieri}}, \bibinfo
  {author} {\bibfnamefont {M.~S.}\ \bibnamefont {Kim}}, \bibinfo {author}
  {\bibfnamefont {V.}~\bibnamefont {Vedral}},\ and\ \bibinfo {author}
  {\bibfnamefont {I.~A.}\ \bibnamefont {Walmsley}},\ }\bibfield  {title}
  {\bibinfo {title} {Photonic {Maxwell's} demon},\ }\href
  {https://doi.org/10.1103/PhysRevLett.116.050401} {\bibfield  {journal}
  {\bibinfo  {journal} {Phys. Rev. Lett.}\ }\textbf {\bibinfo {volume} {116}},\
  \bibinfo {pages} {050401} (\bibinfo {year} {2016})}\BibitemShut {NoStop}%
\bibitem [{\citenamefont {Lu}\ and\ \citenamefont
  {Jarzynski}(2019)}]{e21010065}%
  \BibitemOpen
  \bibfield  {author} {\bibinfo {author} {\bibfnamefont {Z.}~\bibnamefont
  {Lu}}\ and\ \bibinfo {author} {\bibfnamefont {C.}~\bibnamefont {Jarzynski}},\
  }\bibfield  {title} {\bibinfo {title} {A programmable mechanical maxwell’s
  demon},\ }\bibfield  {journal} {\bibinfo  {journal} {Entropy}\ }\textbf
  {\bibinfo {volume} {21}},\ \href {https://doi.org/10.3390/e21010065}
  {10.3390/e21010065} (\bibinfo {year} {2019})\BibitemShut {NoStop}%
\bibitem [{\citenamefont {Leff}\ and\ \citenamefont {Rex}(2014)}]{LeffRex}%
  \BibitemOpen
  \bibfield  {author} {\bibinfo {author} {\bibfnamefont {H.~S.}\ \bibnamefont
  {Leff}}\ and\ \bibinfo {author} {\bibfnamefont {A.~F.}\ \bibnamefont {Rex}},\
  }\href@noop {} {\emph {\bibinfo {title} {Maxwell's Demon: Entropy,
  Information, Computing}}}\ (\bibinfo  {publisher} {Princeton University
  Press},\ \bibinfo {year} {2014})\BibitemShut {NoStop}%
\bibitem [{\citenamefont {Opatrn\'y}\ \emph {et~al.}(2021)\citenamefont
  {Opatrn\'y}, \citenamefont {Misra},\ and\ \citenamefont
  {Kurizki}}]{OpatrnyPRL21}%
  \BibitemOpen
  \bibfield  {author} {\bibinfo {author} {\bibfnamefont {T.}~\bibnamefont
  {Opatrn\'y}}, \bibinfo {author} {\bibfnamefont {A.}~\bibnamefont {Misra}},\
  and\ \bibinfo {author} {\bibfnamefont {G.}~\bibnamefont {Kurizki}},\
  }\bibfield  {title} {\bibinfo {title} {Work generation from thermal noise by
  quantum phase-sensitive observation},\ }\href
  {https://doi.org/10.1103/PhysRevLett.127.040602} {\bibfield  {journal}
  {\bibinfo  {journal} {Phys. Rev. Lett.}\ }\textbf {\bibinfo {volume} {127}},\
  \bibinfo {pages} {040602} (\bibinfo {year} {2021})}\BibitemShut {NoStop}%
\bibitem [{\citenamefont {Brown}\ \emph {et~al.}(2016)\citenamefont {Brown},
  \citenamefont {Friis},\ and\ \citenamefont {Huber}}]{Gau-pas}%
  \BibitemOpen
  \bibfield  {author} {\bibinfo {author} {\bibfnamefont {E.~G.}\ \bibnamefont
  {Brown}}, \bibinfo {author} {\bibfnamefont {N.}~\bibnamefont {Friis}},\ and\
  \bibinfo {author} {\bibfnamefont {M.}~\bibnamefont {Huber}},\ }\bibfield
  {title} {\bibinfo {title} {Passivity and practical work extraction using
  gaussian operations},\ }\href
  {https://doi.org/10.1088/1367-2630/18/11/113028} {\bibfield  {journal}
  {\bibinfo  {journal} {New Journal of Physics}\ }\textbf {\bibinfo {volume}
  {18}},\ \bibinfo {pages} {113028} (\bibinfo {year} {2016})}\BibitemShut
  {NoStop}%
\bibitem [{\citenamefont {Singh}\ \emph {et~al.}(2019)\citenamefont {Singh},
  \citenamefont {Jabbour}, \citenamefont {Van~Herstraeten},\ and\ \citenamefont
  {Cerf}}]{Uttam}%
  \BibitemOpen
  \bibfield  {author} {\bibinfo {author} {\bibfnamefont {U.}~\bibnamefont
  {Singh}}, \bibinfo {author} {\bibfnamefont {M.~G.}\ \bibnamefont {Jabbour}},
  \bibinfo {author} {\bibfnamefont {Z.}~\bibnamefont {Van~Herstraeten}},\ and\
  \bibinfo {author} {\bibfnamefont {N.~J.}\ \bibnamefont {Cerf}},\ }\bibfield
  {title} {\bibinfo {title} {Quantum thermodynamics in a multipartite setting:
  A resource theory of local gaussian work extraction for multimode bosonic
  systems},\ }\href {https://doi.org/10.1103/PhysRevA.100.042104} {\bibfield
  {journal} {\bibinfo  {journal} {Phys. Rev. A}\ }\textbf {\bibinfo {volume}
  {100}},\ \bibinfo {pages} {042104} (\bibinfo {year} {2019})}\BibitemShut
  {NoStop}%
\bibitem [{\citenamefont {Serafini}\ \emph {et~al.}(2020)\citenamefont
  {Serafini}, \citenamefont {Lostaglio}, \citenamefont {Longden}, \citenamefont
  {Shackerley-Bennett}, \citenamefont {Hsieh},\ and\ \citenamefont
  {Adesso}}]{GA-prl}%
  \BibitemOpen
  \bibfield  {author} {\bibinfo {author} {\bibfnamefont {A.}~\bibnamefont
  {Serafini}}, \bibinfo {author} {\bibfnamefont {M.}~\bibnamefont {Lostaglio}},
  \bibinfo {author} {\bibfnamefont {S.}~\bibnamefont {Longden}}, \bibinfo
  {author} {\bibfnamefont {U.}~\bibnamefont {Shackerley-Bennett}}, \bibinfo
  {author} {\bibfnamefont {C.-Y.}\ \bibnamefont {Hsieh}},\ and\ \bibinfo
  {author} {\bibfnamefont {G.}~\bibnamefont {Adesso}},\ }\bibfield  {title}
  {\bibinfo {title} {Gaussian thermal operations and the limits of algorithmic
  cooling},\ }\href {https://doi.org/10.1103/PhysRevLett.124.010602} {\bibfield
   {journal} {\bibinfo  {journal} {Phys. Rev. Lett.}\ }\textbf {\bibinfo
  {volume} {124}},\ \bibinfo {pages} {010602} (\bibinfo {year}
  {2020})}\BibitemShut {NoStop}%
\bibitem [{\citenamefont {Narasimhachar}\ \emph {et~al.}(2021)\citenamefont
  {Narasimhachar}, \citenamefont {Assad}, \citenamefont {Binder}, \citenamefont
  {Thompson}, \citenamefont {Yadin},\ and\ \citenamefont {Gu}}]{Varun}%
  \BibitemOpen
  \bibfield  {author} {\bibinfo {author} {\bibfnamefont {V.}~\bibnamefont
  {Narasimhachar}}, \bibinfo {author} {\bibfnamefont {S.}~\bibnamefont
  {Assad}}, \bibinfo {author} {\bibfnamefont {F.~C.}\ \bibnamefont {Binder}},
  \bibinfo {author} {\bibfnamefont {J.}~\bibnamefont {Thompson}}, \bibinfo
  {author} {\bibfnamefont {B.}~\bibnamefont {Yadin}},\ and\ \bibinfo {author}
  {\bibfnamefont {M.}~\bibnamefont {Gu}},\ }\bibfield  {title} {\bibinfo
  {title} {Thermodynamic resources in continuous-variable quantum systems},\
  }\bibfield  {journal} {\bibinfo  {journal} {npj Quantum Information}\
  }\textbf {\bibinfo {volume} {7}},\ \href
  {https://doi.org/10.1038/s41534-020-00342-6} {10.1038/s41534-020-00342-6}
  (\bibinfo {year} {2021})\BibitemShut {NoStop}%
\bibitem [{\citenamefont {Yadin}\ \emph {et~al.}(2021)\citenamefont {Yadin},
  \citenamefont {Jee}, \citenamefont {Sparaciari}, \citenamefont {Adesso},\
  and\ \citenamefont {Serafini}}]{CGTO}%
  \BibitemOpen
  \bibfield  {author} {\bibinfo {author} {\bibfnamefont {B.}~\bibnamefont
  {Yadin}}, \bibinfo {author} {\bibfnamefont {H.~H.}\ \bibnamefont {Jee}},
  \bibinfo {author} {\bibfnamefont {C.}~\bibnamefont {Sparaciari}}, \bibinfo
  {author} {\bibfnamefont {G.}~\bibnamefont {Adesso}},\ and\ \bibinfo {author}
  {\bibfnamefont {A.}~\bibnamefont {Serafini}},\ }\href@noop {} {\bibinfo
  {title} {Catalytic gaussian thermal operations}} (\bibinfo {year} {2021}),\
  \Eprint {https://arxiv.org/abs/2112.05540} {arXiv:2112.05540 [quant-ph]}
  \BibitemShut {NoStop}%
\bibitem [{\citenamefont {Ding}\ \emph {et~al.}(2017)\citenamefont {Ding},
  \citenamefont {Maslennikov}, \citenamefont {Habl\"utzel},\ and\ \citenamefont
  {Matsukevich}}]{QND-PRL}%
  \BibitemOpen
  \bibfield  {author} {\bibinfo {author} {\bibfnamefont {S.}~\bibnamefont
  {Ding}}, \bibinfo {author} {\bibfnamefont {G.}~\bibnamefont {Maslennikov}},
  \bibinfo {author} {\bibfnamefont {R.}~\bibnamefont {Habl\"utzel}},\ and\
  \bibinfo {author} {\bibfnamefont {D.}~\bibnamefont {Matsukevich}},\
  }\bibfield  {title} {\bibinfo {title} {Cross-kerr nonlinearity for phonon
  counting},\ }\href {https://doi.org/10.1103/PhysRevLett.119.193602}
  {\bibfield  {journal} {\bibinfo  {journal} {Phys. Rev. Lett.}\ }\textbf
  {\bibinfo {volume} {119}},\ \bibinfo {pages} {193602} (\bibinfo {year}
  {2017})}\BibitemShut {NoStop}%
\bibitem [{\citenamefont {Milburn}\ and\ \citenamefont {Walls}(1983)}]{QND1}%
  \BibitemOpen
  \bibfield  {author} {\bibinfo {author} {\bibfnamefont {G.~J.}\ \bibnamefont
  {Milburn}}\ and\ \bibinfo {author} {\bibfnamefont {D.~F.}\ \bibnamefont
  {Walls}},\ }\bibfield  {title} {\bibinfo {title} {Quantum nondemolition
  measurements via quadratic coupling},\ }\href
  {https://doi.org/10.1103/PhysRevA.28.2065} {\bibfield  {journal} {\bibinfo
  {journal} {Phys. Rev. A}\ }\textbf {\bibinfo {volume} {28}},\ \bibinfo
  {pages} {2065} (\bibinfo {year} {1983})}\BibitemShut {NoStop}%
\bibitem [{\citenamefont {Imoto}\ \emph {et~al.}(1985)\citenamefont {Imoto},
  \citenamefont {Haus},\ and\ \citenamefont {Yamamoto}}]{QND2}%
  \BibitemOpen
  \bibfield  {author} {\bibinfo {author} {\bibfnamefont {N.}~\bibnamefont
  {Imoto}}, \bibinfo {author} {\bibfnamefont {H.~A.}\ \bibnamefont {Haus}},\
  and\ \bibinfo {author} {\bibfnamefont {Y.}~\bibnamefont {Yamamoto}},\
  }\bibfield  {title} {\bibinfo {title} {Quantum nondemolition measurement of
  the photon number via the optical kerr effect},\ }\href
  {https://doi.org/10.1103/PhysRevA.32.2287} {\bibfield  {journal} {\bibinfo
  {journal} {Phys. Rev. A}\ }\textbf {\bibinfo {volume} {32}},\ \bibinfo
  {pages} {2287} (\bibinfo {year} {1985})}\BibitemShut {NoStop}%
\bibitem [{\citenamefont {Howell}\ and\ \citenamefont {Yeazell}(2000)}]{QND3}%
  \BibitemOpen
  \bibfield  {author} {\bibinfo {author} {\bibfnamefont {J.~C.}\ \bibnamefont
  {Howell}}\ and\ \bibinfo {author} {\bibfnamefont {J.~A.}\ \bibnamefont
  {Yeazell}},\ }\bibfield  {title} {\bibinfo {title} {Nondestructive
  single-photon trigger},\ }\href {https://doi.org/10.1103/PhysRevA.62.032311}
  {\bibfield  {journal} {\bibinfo  {journal} {Phys. Rev. A}\ }\textbf {\bibinfo
  {volume} {62}},\ \bibinfo {pages} {032311} (\bibinfo {year}
  {2000})}\BibitemShut {NoStop}%
\bibitem [{\citenamefont {Munro}\ \emph {et~al.}(2005)\citenamefont {Munro},
  \citenamefont {Nemoto}, \citenamefont {Beausoleil},\ and\ \citenamefont
  {Spiller}}]{QND4}%
  \BibitemOpen
  \bibfield  {author} {\bibinfo {author} {\bibfnamefont {W.~J.}\ \bibnamefont
  {Munro}}, \bibinfo {author} {\bibfnamefont {K.}~\bibnamefont {Nemoto}},
  \bibinfo {author} {\bibfnamefont {R.~G.}\ \bibnamefont {Beausoleil}},\ and\
  \bibinfo {author} {\bibfnamefont {T.~P.}\ \bibnamefont {Spiller}},\
  }\bibfield  {title} {\bibinfo {title} {High-efficiency quantum-nondemolition
  single-photon-number-resolving detector},\ }\href
  {https://doi.org/10.1103/PhysRevA.71.033819} {\bibfield  {journal} {\bibinfo
  {journal} {Phys. Rev. A}\ }\textbf {\bibinfo {volume} {71}},\ \bibinfo
  {pages} {033819} (\bibinfo {year} {2005})}\BibitemShut {NoStop}%
\bibitem [{\citenamefont {Kol\'a\ifmmode~\check{r}\else \v{r}\fi{}}\ \emph
  {et~al.}(2012)\citenamefont {Kol\'a\ifmmode~\check{r}\else \v{r}\fi{}},
  \citenamefont {Gelbwaser-Klimovsky}, \citenamefont {Alicki},\ and\
  \citenamefont {Kurizki}}]{PhysRevLett.109.090601}%
  \BibitemOpen
  \bibfield  {author} {\bibinfo {author} {\bibfnamefont {M.}~\bibnamefont
  {Kol\'a\ifmmode~\check{r}\else \v{r}\fi{}}}, \bibinfo {author} {\bibfnamefont
  {D.}~\bibnamefont {Gelbwaser-Klimovsky}}, \bibinfo {author} {\bibfnamefont
  {R.}~\bibnamefont {Alicki}},\ and\ \bibinfo {author} {\bibfnamefont
  {G.}~\bibnamefont {Kurizki}},\ }\bibfield  {title} {\bibinfo {title} {Quantum
  bath refrigeration towards absolute zero: Challenging the unattainability
  principle},\ }\href {https://doi.org/10.1103/PhysRevLett.109.090601}
  {\bibfield  {journal} {\bibinfo  {journal} {Phys. Rev. Lett.}\ }\textbf
  {\bibinfo {volume} {109}},\ \bibinfo {pages} {090601} (\bibinfo {year}
  {2012})}\BibitemShut {NoStop}%
\bibitem [{\citenamefont {Gelbwaser-Klimovsky}\ \emph
  {et~al.}(2013{\natexlab{b}})\citenamefont {Gelbwaser-Klimovsky},
  \citenamefont {Alicki},\ and\ \citenamefont {Kurizki}}]{Gelbwaser_2013_a}%
  \BibitemOpen
  \bibfield  {author} {\bibinfo {author} {\bibfnamefont {D.}~\bibnamefont
  {Gelbwaser-Klimovsky}}, \bibinfo {author} {\bibfnamefont {R.}~\bibnamefont
  {Alicki}},\ and\ \bibinfo {author} {\bibfnamefont {G.}~\bibnamefont
  {Kurizki}},\ }\bibfield  {title} {\bibinfo {title} {Minimal universal quantum
  heat machine},\ }\href {https://doi.org/10.1103/PhysRevE.87.012140}
  {\bibfield  {journal} {\bibinfo  {journal} {Phys. Rev. E}\ }\textbf {\bibinfo
  {volume} {87}},\ \bibinfo {pages} {012140} (\bibinfo {year}
  {2013}{\natexlab{b}})}\BibitemShut {NoStop}%
\bibitem [{\citenamefont {Karimi}\ \emph {et~al.}(2017)\citenamefont {Karimi},
  \citenamefont {Pekola}, \citenamefont {Campisi},\ and\ \citenamefont
  {Fazio}}]{Karimi_2017}%
  \BibitemOpen
  \bibfield  {author} {\bibinfo {author} {\bibfnamefont {B.}~\bibnamefont
  {Karimi}}, \bibinfo {author} {\bibfnamefont {J.~P.}\ \bibnamefont {Pekola}},
  \bibinfo {author} {\bibfnamefont {M.}~\bibnamefont {Campisi}},\ and\ \bibinfo
  {author} {\bibfnamefont {R.}~\bibnamefont {Fazio}},\ }\bibfield  {title}
  {\bibinfo {title} {Coupled qubits as a quantum heat switch},\ }\href
  {https://doi.org/10.1088/2058-9565/aa8330} {\bibfield  {journal} {\bibinfo
  {journal} {Quantum Science and Technology}\ }\textbf {\bibinfo {volume}
  {2}},\ \bibinfo {pages} {044007} (\bibinfo {year} {2017})}\BibitemShut
  {NoStop}%
\bibitem [{\citenamefont {Naseem}\ \emph {et~al.}(2020)\citenamefont {Naseem},
  \citenamefont {Misra}, \citenamefont {M\"ustecaplio\ifmmode~\breve{g}\else
  \u{g}\fi{}lu},\ and\ \citenamefont {Kurizki}}]{Tahir}%
  \BibitemOpen
  \bibfield  {author} {\bibinfo {author} {\bibfnamefont {M.~T.}\ \bibnamefont
  {Naseem}}, \bibinfo {author} {\bibfnamefont {A.}~\bibnamefont {Misra}},
  \bibinfo {author} {\bibfnamefont {O.~E.}\ \bibnamefont
  {M\"ustecaplio\ifmmode~\breve{g}\else \u{g}\fi{}lu}},\ and\ \bibinfo {author}
  {\bibfnamefont {G.}~\bibnamefont {Kurizki}},\ }\bibfield  {title} {\bibinfo
  {title} {Minimal quantum heat manager boosted by bath spectral filtering},\
  }\href {https://doi.org/10.1103/PhysRevResearch.2.033285} {\bibfield
  {journal} {\bibinfo  {journal} {Phys. Rev. Research}\ }\textbf {\bibinfo
  {volume} {2}},\ \bibinfo {pages} {033285} (\bibinfo {year}
  {2020})}\BibitemShut {NoStop}%
\bibitem [{\citenamefont {Karg\ifmmode \imath \else~\i \fi{}}\ \emph
  {et~al.}(2019)\citenamefont {Karg\ifmmode \imath \else~\i \fi{}},
  \citenamefont {Naseem}, \citenamefont {Opatrn\'y}, \citenamefont
  {M\"ustecapl\ifmmode \imath \else \i \fi{}o\ifmmode~\breve{g}\else
  \u{g}\fi{}lu},\ and\ \citenamefont {Kurizki}}]{PhysRevE.99.042121}%
  \BibitemOpen
  \bibfield  {author} {\bibinfo {author} {\bibfnamefont {C.}~\bibnamefont
  {Karg\ifmmode \imath \else~\i \fi{}}}, \bibinfo {author} {\bibfnamefont
  {M.~T.}\ \bibnamefont {Naseem}}, \bibinfo {author} {\bibfnamefont {T.~c.~v.}\
  \bibnamefont {Opatrn\'y}}, \bibinfo {author} {\bibfnamefont {O.~E.}\
  \bibnamefont {M\"ustecapl\ifmmode \imath \else \i
  \fi{}o\ifmmode~\breve{g}\else \u{g}\fi{}lu}},\ and\ \bibinfo {author}
  {\bibfnamefont {G.}~\bibnamefont {Kurizki}},\ }\bibfield  {title} {\bibinfo
  {title} {Quantum optical two-atom thermal diode},\ }\href
  {https://doi.org/10.1103/PhysRevE.99.042121} {\bibfield  {journal} {\bibinfo
  {journal} {Phys. Rev. E}\ }\textbf {\bibinfo {volume} {99}},\ \bibinfo
  {pages} {042121} (\bibinfo {year} {2019})}\BibitemShut {NoStop}%
\bibitem [{\citenamefont {Klatzow}\ \emph {et~al.}(2019)\citenamefont
  {Klatzow}, \citenamefont {Becker}, \citenamefont {Ledingham}, \citenamefont
  {Weinzetl}, \citenamefont {Kaczmarek}, \citenamefont {Saunders},
  \citenamefont {Nunn}, \citenamefont {Walmsley}, \citenamefont {Uzdin},\ and\
  \citenamefont {Poem}}]{PhysRevLett.122.110601}%
  \BibitemOpen
  \bibfield  {author} {\bibinfo {author} {\bibfnamefont {J.}~\bibnamefont
  {Klatzow}}, \bibinfo {author} {\bibfnamefont {J.~N.}\ \bibnamefont {Becker}},
  \bibinfo {author} {\bibfnamefont {P.~M.}\ \bibnamefont {Ledingham}}, \bibinfo
  {author} {\bibfnamefont {C.}~\bibnamefont {Weinzetl}}, \bibinfo {author}
  {\bibfnamefont {K.~T.}\ \bibnamefont {Kaczmarek}}, \bibinfo {author}
  {\bibfnamefont {D.~J.}\ \bibnamefont {Saunders}}, \bibinfo {author}
  {\bibfnamefont {J.}~\bibnamefont {Nunn}}, \bibinfo {author} {\bibfnamefont
  {I.~A.}\ \bibnamefont {Walmsley}}, \bibinfo {author} {\bibfnamefont
  {R.}~\bibnamefont {Uzdin}},\ and\ \bibinfo {author} {\bibfnamefont
  {E.}~\bibnamefont {Poem}},\ }\bibfield  {title} {\bibinfo {title}
  {Experimental demonstration of quantum effects in the operation of
  microscopic heat engines},\ }\href
  {https://doi.org/10.1103/PhysRevLett.122.110601} {\bibfield  {journal}
  {\bibinfo  {journal} {Phys. Rev. Lett.}\ }\textbf {\bibinfo {volume} {122}},\
  \bibinfo {pages} {110601} (\bibinfo {year} {2019})}\BibitemShut {NoStop}%
\bibitem [{\citenamefont {Reck}\ \emph {et~al.}(1994)\citenamefont {Reck},
  \citenamefont {Zeilinger}, \citenamefont {Bernstein},\ and\ \citenamefont
  {Bertani}}]{PhysRevLett.73.58}%
  \BibitemOpen
  \bibfield  {author} {\bibinfo {author} {\bibfnamefont {M.}~\bibnamefont
  {Reck}}, \bibinfo {author} {\bibfnamefont {A.}~\bibnamefont {Zeilinger}},
  \bibinfo {author} {\bibfnamefont {H.~J.}\ \bibnamefont {Bernstein}},\ and\
  \bibinfo {author} {\bibfnamefont {P.}~\bibnamefont {Bertani}},\ }\bibfield
  {title} {\bibinfo {title} {Experimental realization of any discrete unitary
  operator},\ }\href {https://doi.org/10.1103/PhysRevLett.73.58} {\bibfield
  {journal} {\bibinfo  {journal} {Phys. Rev. Lett.}\ }\textbf {\bibinfo
  {volume} {73}},\ \bibinfo {pages} {58} (\bibinfo {year} {1994})}\BibitemShut
  {NoStop}%
\bibitem [{\citenamefont {Clements}\ \emph {et~al.}(2016)\citenamefont
  {Clements}, \citenamefont {Humphreys}, \citenamefont {Metcalf}, \citenamefont
  {Kolthammer},\ and\ \citenamefont {Walmsley}}]{Clements:16}%
  \BibitemOpen
  \bibfield  {author} {\bibinfo {author} {\bibfnamefont {W.~R.}\ \bibnamefont
  {Clements}}, \bibinfo {author} {\bibfnamefont {P.~C.}\ \bibnamefont
  {Humphreys}}, \bibinfo {author} {\bibfnamefont {B.~J.}\ \bibnamefont
  {Metcalf}}, \bibinfo {author} {\bibfnamefont {W.~S.}\ \bibnamefont
  {Kolthammer}},\ and\ \bibinfo {author} {\bibfnamefont {I.~A.}\ \bibnamefont
  {Walmsley}},\ }\bibfield  {title} {\bibinfo {title} {Optimal design for
  universal multiport interferometers},\ }\href
  {https://doi.org/10.1364/OPTICA.3.001460} {\bibfield  {journal} {\bibinfo
  {journal} {Optica}\ }\textbf {\bibinfo {volume} {3}},\ \bibinfo {pages}
  {1460} (\bibinfo {year} {2016})}\BibitemShut {NoStop}%
\bibitem [{\citenamefont {Martinez}\ and\ \citenamefont
  {Paz}(2013)}]{PhysRevLett.110.130406}%
  \BibitemOpen
  \bibfield  {author} {\bibinfo {author} {\bibfnamefont {E.~A.}\ \bibnamefont
  {Martinez}}\ and\ \bibinfo {author} {\bibfnamefont {J.~P.}\ \bibnamefont
  {Paz}},\ }\bibfield  {title} {\bibinfo {title} {Dynamics and thermodynamics
  of linear quantum open systems},\ }\href
  {https://doi.org/10.1103/PhysRevLett.110.130406} {\bibfield  {journal}
  {\bibinfo  {journal} {Phys. Rev. Lett.}\ }\textbf {\bibinfo {volume} {110}},\
  \bibinfo {pages} {130406} (\bibinfo {year} {2013})}\BibitemShut {NoStop}%
\bibitem [{\citenamefont {Freitas}\ and\ \citenamefont
  {Paz}(2014)}]{PhysRevE.90.042128}%
  \BibitemOpen
  \bibfield  {author} {\bibinfo {author} {\bibfnamefont {N.}~\bibnamefont
  {Freitas}}\ and\ \bibinfo {author} {\bibfnamefont {J.~P.}\ \bibnamefont
  {Paz}},\ }\bibfield  {title} {\bibinfo {title} {Analytic solution for heat
  flow through a general harmonic network},\ }\href
  {https://doi.org/10.1103/PhysRevE.90.042128} {\bibfield  {journal} {\bibinfo
  {journal} {Phys. Rev. E}\ }\textbf {\bibinfo {volume} {90}},\ \bibinfo
  {pages} {042128} (\bibinfo {year} {2014})}\BibitemShut {NoStop}%
\bibitem [{\citenamefont {Schwinger}(1965)}]{schwinger}%
  \BibitemOpen
  \bibfield  {author} {\bibinfo {author} {\bibfnamefont {J.}~\bibnamefont
  {Schwinger}},\ }\href@noop {} {\emph {\bibinfo {title} {Quantum Theory of
  Angular Momentum: A Collection of Reprints and Original Papers}}},\ edited
  by\ \bibinfo {editor} {\bibfnamefont {H.~V.~D.}\ \bibnamefont
  {L.C.~Biedenharn}}\ (\bibinfo  {publisher} {Academic Press},\ \bibinfo {year}
  {1965})\BibitemShut {NoStop}%
\bibitem [{\citenamefont {Agarwal}(1974)}]{agarwalbook}%
  \BibitemOpen
  \bibfield  {author} {\bibinfo {author} {\bibfnamefont {G.~S.}\ \bibnamefont
  {Agarwal}},\ }\href@noop {} {\emph {\bibinfo {title} {Quantum statistical
  theories of spontaneous emission and their relation to other approaches}}}\
  (\bibinfo  {publisher} {Springer-Verlag},\ \bibinfo {address} {Berlin
  Heidelberg},\ \bibinfo {year} {1974})\BibitemShut {NoStop}%
\bibitem [{\citenamefont {Leonhardt}(1997)}]{Ulf}%
  \BibitemOpen
  \bibfield  {author} {\bibinfo {author} {\bibfnamefont {U.}~\bibnamefont
  {Leonhardt}},\ }\href@noop {} {\emph {\bibinfo {title} {Measuring the Quantum
  State of Light}}}\ (\bibinfo  {publisher} {Cambridge Studies in Modern
  Optics},\ \bibinfo {address} {Cambridge},\ \bibinfo {year}
  {1997})\BibitemShut {NoStop}%
\bibitem [{\citenamefont {Heikkil\"a}\ \emph {et~al.}(2014)\citenamefont
  {Heikkil\"a}, \citenamefont {Massel}, \citenamefont {Tuorila}, \citenamefont
  {Khan},\ and\ \citenamefont {Sillanp\"a\"a}}]{Hekkila2014PRL}%
  \BibitemOpen
  \bibfield  {author} {\bibinfo {author} {\bibfnamefont {T.~T.}\ \bibnamefont
  {Heikkil\"a}}, \bibinfo {author} {\bibfnamefont {F.}~\bibnamefont {Massel}},
  \bibinfo {author} {\bibfnamefont {J.}~\bibnamefont {Tuorila}}, \bibinfo
  {author} {\bibfnamefont {R.}~\bibnamefont {Khan}},\ and\ \bibinfo {author}
  {\bibfnamefont {M.~A.}\ \bibnamefont {Sillanp\"a\"a}},\ }\bibfield  {title}
  {\bibinfo {title} {Enhancing optomechanical coupling via the josephson
  effect},\ }\href {https://doi.org/10.1103/PhysRevLett.112.203603} {\bibfield
  {journal} {\bibinfo  {journal} {Phys. Rev. Lett.}\ }\textbf {\bibinfo
  {volume} {112}},\ \bibinfo {pages} {203603} (\bibinfo {year}
  {2014})}\BibitemShut {NoStop}%
\bibitem [{\citenamefont {Kurizki}\ \emph {et~al.}(2015)\citenamefont
  {Kurizki}, \citenamefont {Bertet}, \citenamefont {Kubo}, \citenamefont
  {Mølmer}, \citenamefont {Petrosyan}, \citenamefont {Rabl},\ and\
  \citenamefont {Schmiedmayer}}]{GKPNAS}%
  \BibitemOpen
  \bibfield  {author} {\bibinfo {author} {\bibfnamefont {G.}~\bibnamefont
  {Kurizki}}, \bibinfo {author} {\bibfnamefont {P.}~\bibnamefont {Bertet}},
  \bibinfo {author} {\bibfnamefont {Y.}~\bibnamefont {Kubo}}, \bibinfo {author}
  {\bibfnamefont {K.}~\bibnamefont {Mølmer}}, \bibinfo {author} {\bibfnamefont
  {D.}~\bibnamefont {Petrosyan}}, \bibinfo {author} {\bibfnamefont
  {P.}~\bibnamefont {Rabl}},\ and\ \bibinfo {author} {\bibfnamefont
  {J.}~\bibnamefont {Schmiedmayer}},\ }\bibfield  {title} {\bibinfo {title}
  {Quantum technologies with hybrid systems},\ }\href
  {https://doi.org/10.1073/pnas.1419326112} {\bibfield  {journal} {\bibinfo
  {journal} {Proceedings of the National Academy of Sciences}\ }\textbf
  {\bibinfo {volume} {112}},\ \bibinfo {pages} {3866} (\bibinfo {year}
  {2015})},\ \Eprint
  {https://arxiv.org/abs/https://www.pnas.org/doi/pdf/10.1073/pnas.1419326112}
  {https://www.pnas.org/doi/pdf/10.1073/pnas.1419326112} \BibitemShut {NoStop}%
\bibitem [{\citenamefont {Shahmoon}\ \emph {et~al.}(2011)\citenamefont
  {Shahmoon}, \citenamefont {Kurizki}, \citenamefont {Fleischhauer},\ and\
  \citenamefont {Petrosyan}}]{Shahmoon2011}%
  \BibitemOpen
  \bibfield  {author} {\bibinfo {author} {\bibfnamefont {E.}~\bibnamefont
  {Shahmoon}}, \bibinfo {author} {\bibfnamefont {G.}~\bibnamefont {Kurizki}},
  \bibinfo {author} {\bibfnamefont {M.}~\bibnamefont {Fleischhauer}},\ and\
  \bibinfo {author} {\bibfnamefont {D.}~\bibnamefont {Petrosyan}},\ }\bibfield
  {title} {\bibinfo {title} {Strongly interacting photons in hollow-core
  waveguides},\ }\href {https://doi.org/10.1103/PhysRevA.83.033806} {\bibfield
  {journal} {\bibinfo  {journal} {Phys. Rev. A}\ }\textbf {\bibinfo {volume}
  {83}},\ \bibinfo {pages} {033806} (\bibinfo {year} {2011})}\BibitemShut
  {NoStop}%
\bibitem [{\citenamefont {Friedler}\ \emph {et~al.}(2005)\citenamefont
  {Friedler}, \citenamefont {Petrosyan}, \citenamefont {Fleischhauer},\ and\
  \citenamefont {Kurizki}}]{Friedler}%
  \BibitemOpen
  \bibfield  {author} {\bibinfo {author} {\bibfnamefont {I.}~\bibnamefont
  {Friedler}}, \bibinfo {author} {\bibfnamefont {D.}~\bibnamefont {Petrosyan}},
  \bibinfo {author} {\bibfnamefont {M.}~\bibnamefont {Fleischhauer}},\ and\
  \bibinfo {author} {\bibfnamefont {G.}~\bibnamefont {Kurizki}},\ }\bibfield
  {title} {\bibinfo {title} {Long-range interactions and entanglement of slow
  single-photon pulses},\ }\href {https://doi.org/10.1103/PhysRevA.72.043803}
  {\bibfield  {journal} {\bibinfo  {journal} {Phys. Rev. A}\ }\textbf {\bibinfo
  {volume} {72}},\ \bibinfo {pages} {043803} (\bibinfo {year}
  {2005})}\BibitemShut {NoStop}%
\bibitem [{\citenamefont {Firstenberg}\ \emph {et~al.}(2013)\citenamefont
  {Firstenberg}, \citenamefont {Peyronel}, \citenamefont {Liang}, \citenamefont
  {Gorshkov}, \citenamefont {Lukin},\ and\ \citenamefont
  {Vuleti{\'{c}}}}]{Firstenberg2013}%
  \BibitemOpen
  \bibfield  {author} {\bibinfo {author} {\bibfnamefont {O.}~\bibnamefont
  {Firstenberg}}, \bibinfo {author} {\bibfnamefont {T.}~\bibnamefont
  {Peyronel}}, \bibinfo {author} {\bibfnamefont {Q.-Y.}\ \bibnamefont {Liang}},
  \bibinfo {author} {\bibfnamefont {A.~V.}\ \bibnamefont {Gorshkov}}, \bibinfo
  {author} {\bibfnamefont {M.~D.}\ \bibnamefont {Lukin}},\ and\ \bibinfo
  {author} {\bibfnamefont {V.}~\bibnamefont {Vuleti{\'{c}}}},\ }\bibfield
  {title} {\bibinfo {title} {Attractive photons in a quantum nonlinear
  medium},\ }\href {https://doi.org/10.1038/nature12512} {\bibfield  {journal}
  {\bibinfo  {journal} {Nature}\ }\textbf {\bibinfo {volume} {502}},\ \bibinfo
  {pages} {71} (\bibinfo {year} {2013})}\BibitemShut {NoStop}%
\bibitem [{\citenamefont {Thompson}\ \emph {et~al.}(2017)\citenamefont
  {Thompson}, \citenamefont {Nicholson}, \citenamefont {Liang}, \citenamefont
  {Cantu}, \citenamefont {Venkatramani}, \citenamefont {Choi}, \citenamefont
  {Fedorov}, \citenamefont {Viscor}, \citenamefont {Pohl}, \citenamefont
  {Lukin},\ and\ \citenamefont {Vuleti{\'{c}}}}]{Thompson2017}%
  \BibitemOpen
  \bibfield  {author} {\bibinfo {author} {\bibfnamefont {J.~D.}\ \bibnamefont
  {Thompson}}, \bibinfo {author} {\bibfnamefont {T.~L.}\ \bibnamefont
  {Nicholson}}, \bibinfo {author} {\bibfnamefont {Q.-Y.}\ \bibnamefont
  {Liang}}, \bibinfo {author} {\bibfnamefont {S.~H.}\ \bibnamefont {Cantu}},
  \bibinfo {author} {\bibfnamefont {A.~V.}\ \bibnamefont {Venkatramani}},
  \bibinfo {author} {\bibfnamefont {S.}~\bibnamefont {Choi}}, \bibinfo {author}
  {\bibfnamefont {I.~A.}\ \bibnamefont {Fedorov}}, \bibinfo {author}
  {\bibfnamefont {D.}~\bibnamefont {Viscor}}, \bibinfo {author} {\bibfnamefont
  {T.}~\bibnamefont {Pohl}}, \bibinfo {author} {\bibfnamefont {M.~D.}\
  \bibnamefont {Lukin}},\ and\ \bibinfo {author} {\bibfnamefont
  {V.}~\bibnamefont {Vuleti{\'{c}}}},\ }\bibfield  {title} {\bibinfo {title}
  {Symmetry-protected collisions between strongly interacting photons},\ }\href
  {https://doi.org/10.1038/nature20823} {\bibfield  {journal} {\bibinfo
  {journal} {Nature}\ }\textbf {\bibinfo {volume} {542}},\ \bibinfo {pages}
  {206} (\bibinfo {year} {2017})}\BibitemShut {NoStop}%
\bibitem [{\citenamefont {Tiarks}\ \emph {et~al.}(2019)\citenamefont {Tiarks},
  \citenamefont {Schmidt-Eberle}, \citenamefont {Stolz}, \citenamefont
  {Rempe},\ and\ \citenamefont {D{\"u}rr}}]{Tiarks2019}%
  \BibitemOpen
  \bibfield  {author} {\bibinfo {author} {\bibfnamefont {D.}~\bibnamefont
  {Tiarks}}, \bibinfo {author} {\bibfnamefont {S.}~\bibnamefont
  {Schmidt-Eberle}}, \bibinfo {author} {\bibfnamefont {T.}~\bibnamefont
  {Stolz}}, \bibinfo {author} {\bibfnamefont {G.}~\bibnamefont {Rempe}},\ and\
  \bibinfo {author} {\bibfnamefont {S.}~\bibnamefont {D{\"u}rr}},\ }\bibfield
  {title} {\bibinfo {title} {A photon--photon quantum gate based on {Rydberg}
  interactions},\ }\href {https://doi.org/10.1038/s41567-018-0313-7} {\bibfield
   {journal} {\bibinfo  {journal} {Nature Physics}\ }\textbf {\bibinfo {volume}
  {15}},\ \bibinfo {pages} {124} (\bibinfo {year} {2019})}\BibitemShut
  {NoStop}%
\bibitem [{\citenamefont {Parrondo}\ \emph {et~al.}(2015)\citenamefont
  {Parrondo}, \citenamefont {Horowitz},\ and\ \citenamefont
  {Sagawa}}]{parrondo2015thermodynamics}%
  \BibitemOpen
  \bibfield  {author} {\bibinfo {author} {\bibfnamefont {J.~M.~R.}\
  \bibnamefont {Parrondo}}, \bibinfo {author} {\bibfnamefont {J.~M.}\
  \bibnamefont {Horowitz}},\ and\ \bibinfo {author} {\bibfnamefont
  {T.}~\bibnamefont {Sagawa}},\ }\bibfield  {title} {\bibinfo {title}
  {Thermodynamics of information},\ }\href {https://doi.org/10.1038/nphys3230}
  {\bibfield  {journal} {\bibinfo  {journal} {Nat. Phys.}\ }\textbf {\bibinfo
  {volume} {11}},\ \bibinfo {pages} {131} (\bibinfo {year} {2015})}\BibitemShut
  {NoStop}%
\bibitem [{\citenamefont {Cover}(1991)}]{Cover}%
  \BibitemOpen
  \bibfield  {author} {\bibinfo {author} {\bibfnamefont {T.~M.}\ \bibnamefont
  {Cover}},\ }\href@noop {} {\emph {\bibinfo {title} {Elements of information
  theory}}}\ (\bibinfo  {publisher} {New York : Wiley},\ \bibinfo {year}
  {1991})\BibitemShut {NoStop}%
\bibitem [{\citenamefont {Misra}\ \emph {et~al.}(2022)\citenamefont {Misra},
  \citenamefont {Opatrný},\ and\ \citenamefont {Kurizki}}]{Misra}%
  \BibitemOpen
  \bibfield  {author} {\bibinfo {author} {\bibfnamefont {A.}~\bibnamefont
  {Misra}}, \bibinfo {author} {\bibfnamefont {T.}~\bibnamefont {Opatrný}},\
  and\ \bibinfo {author} {\bibfnamefont {G.}~\bibnamefont {Kurizki}},\
  }\bibfield  {title} {\bibinfo {title} {Work extraction from single-mode
  thermal noise by measurements: How important is information?},\ }\bibfield
  {journal} {\bibinfo  {journal} {arXiv}\ }\href
  {https://doi.org/10.48550/ARXIV.2201.11567} {10.48550/ARXIV.2201.11567}
  (\bibinfo {year} {2022})\BibitemShut {NoStop}%
\bibitem [{\citenamefont {Boyd}(2008)}]{Boyd}%
  \BibitemOpen
  \bibfield  {author} {\bibinfo {author} {\bibfnamefont {R.~W.}\ \bibnamefont
  {Boyd}},\ }\href
  {https://doi.org/https://doi.org/10.1016/B978-0-12-369470-6.00020-4} {\emph
  {\bibinfo {title} {Nonlinear Optics}}}\ (\bibinfo  {publisher} {Academic
  Press},\ \bibinfo {address} {Burlington},\ \bibinfo {year}
  {2008})\BibitemShut {NoStop}%
\bibitem [{\citenamefont {Jauch}\ and\ \citenamefont {Rohrlich}(1976)}]{Jauch}%
  \BibitemOpen
  \bibfield  {author} {\bibinfo {author} {\bibfnamefont {J.~M.}\ \bibnamefont
  {Jauch}}\ and\ \bibinfo {author} {\bibfnamefont {F.}~\bibnamefont
  {Rohrlich}},\ }\href@noop {} {\emph {\bibinfo {title} {The Theory of Photons
  and Electrons}}}\ (\bibinfo  {publisher} {Springer},\ \bibinfo {year}
  {1976})\BibitemShut {NoStop}%
\bibitem [{\citenamefont {Robson}(1974)}]{Robson}%
  \BibitemOpen
  \bibfield  {author} {\bibinfo {author} {\bibfnamefont {B.~A.}\ \bibnamefont
  {Robson}},\ }\href@noop {} {\emph {\bibinfo {title} {The Theory of
  Polarization Phenomena}}}\ (\bibinfo  {publisher} {Oxford},\ \bibinfo {year}
  {1974})\BibitemShut {NoStop}%
\bibitem [{\citenamefont {Jaynes}(1957)}]{PhysRev.106.620}%
  \BibitemOpen
  \bibfield  {author} {\bibinfo {author} {\bibfnamefont {E.~T.}\ \bibnamefont
  {Jaynes}},\ }\bibfield  {title} {\bibinfo {title} {Information theory and
  statistical mechanics},\ }\href {https://doi.org/10.1103/PhysRev.106.620}
  {\bibfield  {journal} {\bibinfo  {journal} {Phys. Rev.}\ }\textbf {\bibinfo
  {volume} {106}},\ \bibinfo {pages} {620} (\bibinfo {year}
  {1957})}\BibitemShut {NoStop}%
\bibitem [{\citenamefont {Aspelmeyer}\ \emph {et~al.}(2014)\citenamefont
  {Aspelmeyer}, \citenamefont {Kippenberg},\ and\ \citenamefont
  {Marquardt}}]{RevModPhys.86.1391}%
  \BibitemOpen
  \bibfield  {author} {\bibinfo {author} {\bibfnamefont {M.}~\bibnamefont
  {Aspelmeyer}}, \bibinfo {author} {\bibfnamefont {T.~J.}\ \bibnamefont
  {Kippenberg}},\ and\ \bibinfo {author} {\bibfnamefont {F.}~\bibnamefont
  {Marquardt}},\ }\bibfield  {title} {\bibinfo {title} {Cavity optomechanics},\
  }\href {https://doi.org/10.1103/RevModPhys.86.1391} {\bibfield  {journal}
  {\bibinfo  {journal} {Rev. Mod. Phys.}\ }\textbf {\bibinfo {volume} {86}},\
  \bibinfo {pages} {1391} (\bibinfo {year} {2014})}\BibitemShut {NoStop}%
\bibitem [{\citenamefont {Lloyd}\ and\ \citenamefont
  {Braunstein}(1999)}]{PhysRevLett.82.1784}%
  \BibitemOpen
  \bibfield  {author} {\bibinfo {author} {\bibfnamefont {S.}~\bibnamefont
  {Lloyd}}\ and\ \bibinfo {author} {\bibfnamefont {S.~L.}\ \bibnamefont
  {Braunstein}},\ }\bibfield  {title} {\bibinfo {title} {Quantum computation
  over continuous variables},\ }\href
  {https://doi.org/10.1103/PhysRevLett.82.1784} {\bibfield  {journal} {\bibinfo
   {journal} {Phys. Rev. Lett.}\ }\textbf {\bibinfo {volume} {82}},\ \bibinfo
  {pages} {1784} (\bibinfo {year} {1999})}\BibitemShut {NoStop}%
\end{thebibliography}%

\end{document}